\documentclass[11pt]{article}
\usepackage[a4paper,hmarginratio=1:1,vmarginratio=2:3,totalwidth=15.6cm,
totalheight=21.1cm]{geometry}
\usepackage{bm,epstopdf,epsfig,amsmath,amssymb,amsfonts,latexsym,comment}
\usepackage[rflt]{floatflt}
\usepackage[font=md,captionskip=8pt]{subfig}
\renewcommand{\baselinestretch}{1.38}

\newcommand{\hc}{\mbox{h.c.}}
\newcommand{\Dslash}{\ensuremath 
\raisebox{0.025cm}{\slash}\hspace{-0.32cm} D}
\newcounter{oldcounter}
\addtocounter{equation}{1}
\begin{document}
 \begin{flushright}
OUTP-09-06P\\
\end{flushright}
\thispagestyle{empty}

\vspace{2.5cm}
\begin{center}
{\Large {\bf  Electroweak and Dark Matter 
Constraints on a $Z^\prime$ \\

\medskip
in Models with a Hidden Valley}}
\vspace{1.cm}

  {\bf S.~Cassel \footnote{\,e-mail: s.cassel1@physics.ox.ac.uk}
  ,\,  D.~M.~Ghilencea \footnote{\,e-mail: d.ghilencea1@physics.ox.ac.uk}
  ,\,  G.~G.~Ross \footnote{\,e-mail: g.ross1@physics.ox.ac.uk}
  }\\
 
\vspace{0.5cm}
 {Rudolf Peierls Centre for Theoretical Physics, University of Oxford,\\
 1 Keble Road, Oxford OX1 3NP, United Kingdom.}
 \end{center}

\medskip

\begin{abstract}
\noindent
We consider current precision electroweak data, $Z^\prime$ searches
 and dark matter constraints and analyse their implications for 
an extension of the SM that includes an extra $U(1)^\prime$ massive 
gauge boson and a particular hidden sector (``hidden valley'') with 
a confining (QCD-like) gauge group.
The constraints on the $Z^\prime$ with arbitrary $Z-Z^\prime$ kinetic
 mixing coming from direct searches and precision tests of the
 Standard Model are analysed and shown to lead to a lower limit of
 $800$ GeV on its mass.
Renormalisable interactions involving the  $Z^\prime$ probe
 the physics of  the hidden valley sector which contains  
a pseudoscalar dark matter candidate. We find that dark matter 
constraints place an upper bound  on the mass of the $Z'$ 
of  $O(10)$ TeV. A TeV mass scale is needed 
for the hidden valley states, and the Sommerfeld factor 
for p-wave dark matter annihilation is found significantly to 
suppress  the allowed parameter space of the model.
\end{abstract}

\bigskip
\bigskip

\def\thefootnote{\arabic{footnote}}
\setcounter{footnote}{0}
\newpage

\setcounter{page}{1}
\tableofcontents{} 
\section{Introduction}

 The Standard Model (SM) has so far proved remarkably successful. With the 
 Large Hadron Collider (LHC) about to start operating, confirmation
 of its Higgs sector  or searches for new  physics beyond
 the SM enter a new, interesting stage, when theoretical
 ideas and  models advocated over the past three decades face the  
 test of new experimental data.
 In this paper we consider current precision 
 electroweak data and dark matter constraints and analyse their
 implications for a rather simple extension of the SM. 
 This consists of  the SM  and a particular hidden sector 
 that contains matter which interacts under a new confining
 gauge group, with an extra $U(1)'$ massive boson that couples to
 SM and hidden sector matter.
 It was partly motivated by novel LHC signatures and 
 search strategies specific to
 this type of model.  Further interest in this class of models
 has also been triggered by 
 attempts to explain the excesses in the positron spectrum measured 
 by PAMELA~\cite{Adriani:2008zr} and ATIC~\cite{Chang:2008zzr}.

Hidden sectors are a common presence in model building beyond the
SM, in models with branes in extra dimensions and in
string  and supersymmetric models, with many implications such as, 
for  example, supersymmetry breaking \cite{Nilles:1983ge}.
 The physics of the hidden sector 
can in principle be probed by the mediating particles communicating 
between the hidden and visible sectors.  While gravitational
 interactions are the usual standard mediators, renormalisable operators 
involving messenger  fields such as new gauge bosons, 
scalars, neutralinos or messenger fermions may  
also be present. For this latter case, a special class of hidden
 sectors  is that advocated in \cite{s1},  also called "hidden valley"
 (HV),  where the hidden sector contains some light states in 
comparison to  the energy scale of the mediator. 
If these light states are able to 
decay to SM states, the dynamics of the HV models can then produce 
some novel phenomenology \cite{s1,s2,s3,Han:2007ae,s5,s6}, possibly 
relevant to the LHC. To select search strategies for such 
new physics, the restrictions that exist on these models from
 electroweak and dark matter constraints need to be 
well understood, and this is the main task of this paper.

In this paper we present the constraints on a ``hidden valley"
extension of the SM with a new neutral gauge boson ($Z^\prime$) 
dominantly responsible for mediation 
between the SM and HV sectors. In Section~\ref{HVM}
we present the hidden valley model 
considered, based on the construction of Strassler \cite{s1},
and define the parameters 
that will be constrained by current experimental data. In this model the HV
matter is confined by a strong (QCD-like) interaction in the hidden sector. 
The physics of the confining theory is  discussed and the spectrum 
and stability of the HV states determined. In Section~\ref{CZ'}
 we review the  constraints on the 
$Z^\prime$  from electroweak precision data (EWPD) and direct searches,
 leading to a {\sl lower mass bound} for the mediator.  The discussion
 and conclusions of this section are more general and 
independent of the physics  of the hidden sector.
The numerical limits presented allow for the possibility of arbitrarily 
large kinetic mixing between the hypercharge and $U(1)^\prime$ gauge 
bosons. So far as we are aware, these are new results 
and apply to the case  that the SM matter have $O(1)$ $U(1)^\prime$ charges.
In Section~\ref{dmc}, the consequences of a dark matter candidate in the hidden
sector are considered, leading to constraints on the HV dark matter candidate. 
This will produce an {\sl upper mass bound} on the $Z^\prime$ mediator. 
We demonstrate that the lower and upper mass bounds 
can be simultaneously satisfied in a finite region of parameter space
that we identify in detail.

The Appendix provides some calculations and results used in the text.
In Appendix~\ref{appendix0} additional details on neutrinos masses
are provided, while in Appendix~\ref{appendixA}
the gauge boson eigenstates are found after the addition of the
local $U(1)^\prime$ symmetry to the SM.  In Appendix~\ref{appendixB}
we list the perturbative gauge boson partial decay rates and
 spin-0 particle annihilation cross sections 
used in Section~\ref{dmc}. 
In Appendix~\ref{appendixC}, we discuss non-perturbative corrections to the 
matrix element for slow moving particles, and present the Sommerfeld factor 
for p-wave multiparticle states in the presence of Yukawa interactions used 
for the dark matter relic density calculations.

\section{A ``hidden valley'' extension of the Standard Model}\label{HVM}

Here we construct a simple extension of the SM 
based on the construction in \cite{s1},
which can lead to an interesting non-standard phenomenology.
The choice of hidden valley matter content and symmetries is 
somewhat arbitrary.  In addition to the non-supersymmetric 
SM particle content with right-handed neutrinos 
($q_i ,u_i,d_i,l_i^{},e_i^{},N_i^{},H$), four new Weyl fermions
($U_{L,R}^{}, D_{L,R}^{}$) are introduced that make up the
hidden sector and a new scalar $\phi$ is added. The new states
are singlets under the SM gauge group $G_{\mbox{\tiny SM}}^{}$.
An extra $U(1)'$ gauge group is also assumed under 
which both SM fields and hidden sector ones are charged, 
and so by construction the associated gauge boson mediates 
between the SM and HV sectors. 

\vspace{2mm}
\begin{table}[h]
\center
\begin{tabular}[c]{||*{6}{c|}|c|c||*{4}{c|}|}\hline
\multicolumn{6}{||c||}{SM fermions} & 
\multicolumn{2}{|c||}{Scalars} & 
\multicolumn{4}{|c||}{HV fermions}
\\ \hline \hline
$q_i^{}$ &$u_i^{ c}$& $ d_i^{ c}$ 
& $\ell_i$ & $e^{c}_i$ & $N_i^{c}$ 
& $H$ & $\phi$
& $U_L^{}$ & $ U_R^{\, c}$ & $D_L^{}$ & $ D_R^{\, c}$   
\\ \hline 
$-\frac15$ & $-\frac15$  & $\frac35$ 
& $\frac35$ & $-\frac15$ & $-1$ 
& $\frac{2}{5}$ & $Q^\prime_\phi$
& $q_+$ & $q_-$ & $-q_+$ & $-q_-$ 
\\ \hline
\end{tabular}
\caption{Charges of the fields under the $U(1)^\prime$
 that extends $G_{\mbox{\tiny SM}}^{}$. }
\label{charges1}
\end{table}

\noindent
A simple way to ensure the cancellation of anomalies 
associated with the $U(1)^\prime$  is to choose
the $U(1)^\prime$ charges for the SM fields to be those of the
family independent $U(1)_\chi^{}$ subgroup of a $SO(10)$ GUT, as
shown in Table~\ref{charges1}.
As discussed in Appendix~\ref{appendix0}
this model may be considered a low energy realisation of a
 theory with $SO(10)$ symmetry.
However, we do not restrict the analysis to be consistent with $SO(10)$ 
unification, but this may be recovered as a special case. It 
follows with the particle content and $U(1)^\prime$ charge 
assignment in Table~\ref{charges1} that the ``un-unified" HV 
model is still gauge anomaly free. Other $U(1)^\prime$ extensions 
are possible and will bring different numerical limits on the model 
parameters compared to those presented in sections~\ref{CZ'} and~\ref{dmc}.
However, as we shall see, the qualitative features of the allowed 
parameter space found are universal. 
The vacuum expectation value (vev) of $\phi$ is responsible for
generating a mass for the $Z'$ which we will find is of $O(TeV)$.
To allow for  right handed neutrinos  masses we choose $Q_\phi'=2$.

One constraint on the hidden valley model which must be satisfied 
is the upper bound on the number of
relativistic degrees of freedom in thermal equilibrium at the 
time of Big Bang 
Nucleosynthesis (BBN). This can be satisfied by ensuring that there 
is a sufficient mass gap for the new states that this model
 introduces. In the hidden sector, interactions other than $U(1)'$
 that are restricted to this sector can also exist. We assume 
as in \cite{s1} that there is a confining $SU(N_v^{})$ QCD-like 
interaction with the 
HV  fermions in the fundamental representation and with a confinement 
scale $\Lambda_{\mbox{\tiny HV}}^{}$. 
If $\Lambda_{\mbox{\tiny HV}}^{}\! >\! \Lambda_{\mbox{\tiny QCD}}^{}$, 
a sufficient mass gap would be provided to satisfy the BBN constraint. 
This hierarchy can naturally set the mass scale of the HV 
states to be significantly different from that of SM states, 
using arguments similar to those in technicolour models. 
The spectrum of  bound states is discussed in  Section~\ref{qcdlike}.

To allow current masses 
for the HV fermions via the Higgs mechanism, the condition 
$q_+ + q_- = -Q_\phi^\prime$ must be satisfied. 
The field $\phi$ can then also act as a 
mediator between the SM and HV sector, via  interactions with the
SM Higgs of the form $|H|^2 \, |\phi|^2$. This term produces mass mixing
terms between the scalars as well, after  breaking  the $U(1)^\prime$ and
electroweak symmetry. The use of a scalar mediator
 (also known as the ``Higgs portal'') has received recent attention in 
 \cite{Wilczek:2004uy,Schabinger:2005ei,w1,hjkl1,j1}.  
In this paper we expect  that mediation via the Higgs portal 
is negligible due to small Yukawa couplings, and the large mass
for the $\phi$, with the resultant small Higgs-$\phi$ mixing, 
making the $Z^\prime$ interactions dominant. 
Indeed the Higgs portal can be ignored completely in this model
if the HV fermions current masses are entirely generated from instanton
 effects after spontaneous chiral symmetry breaking 
\cite{Caldi:1977rc}. Alternatively, if $q_+ + q_- = 0$, mass terms 
for the HV fermions are unprotected by gauge symmetries and they do 
not couple directly to the fundamental scalar fields.

\subsection{$U(1)'$ physics.}\label{U(1)}

The addition of the $U(1)^\prime$ gauge boson to the SM introduces 
interesting physics by itself \cite{Babu:1996vt,zpr1}. The
discussion here, and in Section~\ref{CZ'} for the limits on the
$Z^\prime$ from corrections to SM processes, is 
independent of the details of the hidden sector physics. 
In the most general case, the Lagrangian can contain a term that 
leads to kinetic mixing between the Abelian gauge fields 
$U(1)_Y^{}$ and $U(1)^\prime$:
\medskip
\begin{eqnarray}
{\cal{L}}_{\mbox{\tiny gauge kinetic}}^{}  
=
-\frac14 \left( F_Y^{\mu\nu}  F_{Y\,\mu\nu} ^{}
+ F_{}^{\prime\,\mu\nu} \,F_{\mu\nu}^{\prime}
 + 2 \sin \chi \, F_{\mu\nu}^{\prime} 
F_Y^{\mu\nu} \right) 
+ \mbox{non-Abelian kinetic terms}.
\end{eqnarray}

\medskip\noindent
The presence of kinetic mixing and mass mixing terms changes the 
currents coupling to the gauge fields from that expected in the SM. 
Since the new scalar, $\phi$, is a singlet under $G_{\mbox{\tiny SM}}^{}$, the 
only unbroken Abelian symmetry after $U(1)^\prime$ and electroweak 
symmetry breaking (EWSB) is 
 that associated with electric charge conservation, 
generating the usual structure for electromagnetism. 
Thus, the physical photon has the same couplings as in the SM, and there is no
milli-(electrically)-charged matter in this model at the tree level. 
After EWSB, an $F^\prime_{\mu\nu} F_{\mbox{\tiny EM}}^{\mu\nu}$ term is 
generated by loops. For $U(1)^\prime$ couplings in the
perturbative regime, this term is negligible for the following analysis.

The currents $J_{A,Z,Z^\prime}^{}$, which include the relevant 
coupling constants, associated with the canonically normalised 
gauge boson mass and kinetic eigenstates of the model are related to SM and
 $U(1)^\prime$ currents $J_{\mbox{\tiny EM,Z}}^{\mbox{\tiny SM}}, J^\prime$
as follows (see appendix~A for details):
\vspace{1mm}
\begin{eqnarray}
\left( \begin{array}{c} J_{A }^{}  \\  J_{Z }^{} \\  J_{Z^{\prime}
  }^{} \end{array}\right)
=
\left( \begin{array}{ccc} 1 & 0 & 0 \\  - \cos \theta_w
 \tan \chi \sin \zeta & \,\,\,\,\,  \sin \theta_w \tan \chi \sin \zeta
  + \cos \zeta  \,\,\,\,\, & \sec \chi \sin \zeta \\ - \cos \theta_w
  \tan \chi \cos \zeta & \,\, \sin \theta_w \tan \chi \cos \zeta -
  \sin \zeta  \,\, &  \sec \chi
 \cos \zeta \end{array}\right)
\left( \begin{array}{c} J_{\mbox{\tiny EM} }^{\mbox{\tiny SM} }  \\  
J_Z^{\mbox{\tiny SM} }  \\  
J_{}^{\prime}
\end{array}\right)
\label{jtrans}
\end{eqnarray}
\begin{eqnarray}
\tan (2  \zeta) &=&   \frac{2 \Delta \, (m^2_{Z^\prime} - m^2_{W}
 \sec^2 \theta_w )}{  ( m^2_{Z^\prime} - m^2_{W} \sec^2 \theta_w
 )^2 - \Delta^2 } \\
\nonumber \\
\Delta &=&
\frac{m^{2}_W \,  \sin \theta_w}{\cos^2 \theta_w  \cos \chi}
 \left(   \frac{2\ \! g^\prime \, Q_H^\prime }{g^{}_Y}   - \sin \chi  \right) 
\label{Delta}
\end{eqnarray}
The $\zeta$ angle accounts for the rotation necessary to diagonalise
 the mass matrix, after kinetic mixing has been removed by a field 
redefinition. Note that if $\zeta = 0$, the coupling to the physical 
$Z$ is identical to the SM case, but the $Z^\prime$ coupling still
 varies with the kinetic mixing $\chi$. It is therefore not necessary
 for any SM particles to be charged under the $U(1)^\prime$ for the physical
 $Z^\prime$ to mediate between the SM and HV sectors as this could
 be achieved by the presence of kinetic mixing alone. In
 eq~(\ref{jtrans}), the Weinberg angle is defined as
 $\theta_w^{} \equiv \arctan (g_Y^{} /g_W^{} )$, and  $Q_H^\prime$
 in eq~(\ref{Delta}) is the $U(1)^\prime$ charge of the SM Higgs field.
The tree-level mass of the physical $Z$ deviates from the SM prediction:
\vspace{0mm}
\begin{eqnarray}
m_Z^2 &=& \frac{2 \, m_W^2\, \sec 2\zeta
 \, \sec^2 \theta_w + m_{Z^\prime}^2 (1-\sec 2\zeta)}{1+\sec 2\zeta}
\label{zmass}
\end{eqnarray}
The presence of a local $U(1)^\prime$ symmetry thus affects ``low
energy" electroweak physics whose parameters have been measured very
accurately. The impact of experimental measurements  
on the mixing parameters, $Z'$ mass, $U(1)^\prime$ coupling constant 
and matter charges are analysed in detail in Section~\ref{CZ'}.

\subsection{Physics of a QCD-like interaction in the hidden valley}
\label{qcdlike}

 Let us now address the details of the physics of the hidden sector.
 We consider the case that the HV matter is confined by the
 strong dynamics of an unbroken gauge symmetry in the hidden sector.
 We review the behaviour of such QCD-like theories where the elementary
 matter fields are vector-like pairs of fermions and note some possible
 differences with SM QCD phenomenology. Here we  focus on the lightest
 composite states since, if stable, these could be the dominant dark
 matter candidates. The experimental constraints on the hidden valley
 model with a dark matter candidate are discussed in Section~\ref{dmc}.

 The Lagrangian with elementary fermions in the fundamental
 representation of a confining $SU(N_v^{})$ symmetry ($N_v^{} > 2$)
 includes the terms:
\begin{eqnarray}
{\cal{L}_{\mbox{\tiny HV gauge}}^{} } &=& 
-\frac14 \, {\cal G}_{\mu \nu}^{c}  {\cal G}^{c \mu \nu} -
 \frac{g^2 \, \theta_{\mbox{\tiny HV}}^{}  }{32\, \pi^2} ~ {\cal
   G}_{\mu \nu}^{c}
  \tilde{{\cal G}}^{c \mu \nu} 
\label{hvg} 
\\[9pt]
{\cal{L}_{\mbox{\tiny HV fermion}}^{} } &=& 
\bar{\Psi}_L^{} \,  i  ~ \Dslash \, \Psi_L^{} + \bar{\Psi}_R^{}  \, 
i ~ \Dslash \, \Psi_R^{}
- \bar{\Psi}_R^{} \, M \,  \Psi_L^{} - \bar{\Psi}_L^{} \, M^\dagger   
\, \Psi_R^{}
\label{hvf}
\end{eqnarray}
 where ${\cal G}$ is the field strength of the HV-confined gauge field,
 and chiral multiplets ($\Psi$) have been formed out of the left- and
 right-handed Weyl  fermions ($U,D$). Axial rotations can be used to
 redefine the valley-quark (or ``v-quark") fields, which allows phases to be
 removed from the v-quark mass matrix. However, a $U(1)_A^{}$
 transformation generates non-zero surface terms which contribute to 
the ${\cal G} \tilde{{\cal G}}$ term \cite{'tHooft:1986nc}. The effect 
of choosing a basis where the transformed mass matrix, $M^\prime$, is 
real and positive semi-definite, is to shift the 
$\theta_{\mbox{\tiny HV}}^{}$ parameter:
\begin{eqnarray}
\theta_{\mbox{\tiny HV}}\, \rightarrow ~
 \theta_{\mbox{\tiny HV}}^{} + \mbox{arg} \,
 \det M ~~ \equiv \,\bar{\theta}_{\mbox{\tiny HV}}^{} 
\label{thetsh}
\end{eqnarray}
While in the SM sector, $|\bar{\theta}_{\mbox{\tiny QCD}}^{}| \lesssim
10^{-9}$, in the present case nothing prevents 
$\bar\theta_{\mbox{\tiny HV}}^{}$ from
taking values of order  $O(1)$. This term explicitly violates CP symmetry 
unless $\sin \bar{\theta} = 0$.

The dynamics of the confined matter
can be modelled using an effective theory built to possess the
conserved and approximate symmetries, as detailed in the following.
The kinetic terms of ${\cal{L}_{\mbox{\tiny HV fermion}}^{} }$ are
invariant under independent transformations of the left- and
right-handed 
multiplets. Thus, there is a global $U(N^{})_L^{} \times U(N^{})_R^{}$ 
symmetry, where $N$ is the number of HV fermion flavours. The mass
terms explicitly break the chiral symmetries down to a vector
subgroup. If all flavours have a common current mass, the final
symmetry is $U(N)_V^{}$. If the current masses are not degenerate, 
the final symmetry is $\left[U(1)_V^{}\right]^{N}$.

If there were no fermion masses, the axial symmetry would be
spontaneously broken by the v-quark condensate whilst still
preserving the vector symmetries. Thus, if there are $N^\prime$
fermions whose current masses are much smaller than the confinement
scale, then it is a good approximation to assume that the chiral
symmetry of the kinetic term is first explicitly broken to
$U(N^\prime)_L^{} \times U(N^\prime)_R^{} 
\times \left[U(1)_V^{}\right]^{(N - N^\prime)}$, and then
spontaneously
broken to $U(N^\prime)_V^{} \times \left[U(1 )_V^{}\right]^{(N -
 N^\prime)}$.
The strong interaction is flavour blind in the limit of zero current
masses, so does not spontaneously break the vector symmetry further.
The pseudo-Goldstone bosons associated with the spontaneous
breaking of the approximate axial symmetry are the analogues
of the pions, kaons, eta and eta primed mesons in QCD.

It is convenient to use chiral perturbation theory which is an
expansion in terms of the explicit symmetry breaking parameter,
$M/\Lambda_{\mbox{\tiny HV}}^{}$, to model the pseudo-Goldstone 
boson composite states. The degrees of freedom are parametrised 
by a field matrix, $\Sigma$, which is a function of the 
$N^{\prime 2}$ pseudo-Goldstone bosons:
\begin{eqnarray}
\big( \Psi_L^{} \big)_i \big( \bar{\Psi}_R^{} \big)_j^{}
&=&
-v^3 \, \Sigma_{ij}^{}
\end{eqnarray}
where $v$ accounts for the magnitude of the quark condensate and
 has mass dimension one, and $\Sigma$ is an $N^\prime \times N^\prime$
 matrix. Under a $U(N^\prime)_L^{} \times U(N^\prime)_R^{}$
 transformation, the fields transform as follows:
\begin{eqnarray}
\Psi_L^\prime &=& L \, \Psi_L^{} \nonumber\\
\Psi_R^\prime &=& R \, \Psi_R^{} \nonumber\\
\Sigma^\prime &=& L \Sigma R^\dagger 
\end{eqnarray}

\noindent
An exponential representation for $\Sigma$ is chosen here,
 defined using the decay constant, $f_\pi^{}$ \!:
\begin{eqnarray}\label{udef} 
\Sigma &=& e^{\, 2 i \Pi / f_\pi^{} },
\qquad
\Pi \equiv    \frac{ \eta^\prime   }{\sqrt{2 N^\prime }}
 \, \mathbf{1}     +   \pi_a^{} \, \mathbf{T}_a^{} \, . 
\end{eqnarray}
The $T_a^{}$ generators are normalised so that
$\mbox{Tr}[T_a^{} T_b^{}]~=~\delta_{ab}^{}/2$. 
The pseudo-Goldstone bosons found after spontaneous axial symmetry
breaking are pseudoscalars. The effective Lagrangian for
$\Sigma$, to leading order in $M$ and momentum, is given
 by  \cite{Manohar}:
\vspace{1mm}
\begin{eqnarray}
{\cal{L} } &=& \frac{f_\pi^2}{4} \left[  \mbox{Tr} \left[ D^\mu
    \Sigma^\dagger D_\mu^{} \Sigma \right] +   \Big(  2 \mu \,
  \mbox{Tr} \left[ \Sigma M \right]   + \mbox{h.c} \Big)  +
  \frac{1}{N^\prime} \left( \frac{f_{\eta^\prime}^2 }{f_\pi^2}
 -1 \right) \left| \mbox{Tr} \,[ \Sigma^\dagger  D_\mu^{} \Sigma ]
 \right|^2 \right]    
\label{lcpt}
\end{eqnarray} 
where $\mu = 2 v^3 / f_\pi^2$. Although the axial $U(1)$ symmetry
 ($L = R^\dagger$) of the kinetic terms is classically conserved
 in the limit $M \to 0$, it is always broken by the triangle anomaly
 at the loop level. 
The effect of the anomaly in a $SU(N_v^{})$ confining theory is
 vanishing in the limit $N_v^{} \to \infty$.
 To describe physics at finite $N_v^{}$, perturbations can be added with
 increasing powers of $1/N_v^{}$ \cite{Di Vecchia:1980ve}. This is
 important for correctly describing the $\eta^\prime$ pseudo-Goldstone boson.

The triangle anomaly also controls the ${\cal G} \tilde{{\cal G}}$ term. It
 has been shown by 't Hooft \cite{'tHooft:1986nc} that this term can
 be rewritten as a determinantal interaction of the quark fields,
 $\det [q \bar{q}]$. The leading order correction in $1/N_v^{}$ to the
 effective Lagrangian, which explicitly
 breaks the $U(1)_A^{}$ symmetry, is then given below:
\begin{eqnarray}
\delta_a^{} {\cal{L} } &\propto&  e^{i \, \theta_{\mbox{\tiny HV}}^{}
 } \det [\Sigma]  + \hc
\label{1on}
\end{eqnarray}
The vacuum alignment of the quark condensate in the flavour space,
$\langle \Sigma \rangle$, is selected where the potential of the
effective theory is minimised. Once this is found, the fields can be
expanded about their vevs to determine physical parameters such as
the masses and couplings.

 Before proceeding further, along this line, let us
consider the nature of the dark matter candidates.
 As discussed later in  Section~\ref{DMC}, 
 the flavoured ``valley-pions" (or ``v-pions,"
 $\pi_v^{}$) are the dark matter candidates, which is
 possible since a flavour symmetry can protect them from decay.
 The pion analogues are the
 lightest states in this effective theory, with the $\pi_v^0 ~(=
 \pi_3^{})$ being flavour neutral, and $\pi_v^\pm ~(\propto \pi_1^{}
 \mp i \pi_2^{})$ flavoured. The $\pm,0$ indices of the v-pions
 refer to their flavour isospin as they are electrically neutral. It
 will be shown that the dominant annihilation mechanism of the stable
 v-pions is via the $Z^\prime$ when $\Lambda^{}_{\mbox{\tiny
 HV}}  \gtrsim 30 \, m^{}_{\pi^{}_v}$ or if there is a 
 fractional mass splitting
 of $O(10\%)$ between the (heavier) $\pi_v^0$ and the $\pi_v^\pm$.
 In the remainder of this section we
 show that such a mass  splitting is indeed possible.

 To first order in $M$, the v-pion masses are degenerate
 (neglecting any mass mixing effects):
\begin{eqnarray}
m_{vp}^2 &=& \mu  \sqrt{ m_a^2 + m_b^2 + 2 \, m_a^{} m_b^{} \cos
  \bar{\theta}_{\mbox{\tiny HV}}^{} } 
\label{mvp}
\end{eqnarray}
where $m_{a,b}^{}$ are the magnitudes of the HV fermion current 
masses. This degeneracy is lifted by $O(M^2)$ terms\footnote{The
 degeneracy is also lifted if the bosons interact non-universally with 
other fields.}:
\begin{eqnarray}
\delta_m^{} {\cal{L}} &=&
\frac{f_\pi^2}{4} \left[ \left( c_1^{} \mbox{Tr} [\Sigma M]^2 +
 c_3^{} \mbox{Tr} [\Sigma M\Sigma M] + \mbox{h.c.} \right) +2 c_2^{}
 \left| \mbox{Tr} [\Sigma M] \right|^2 \right]  .
\label{hgo}
\end{eqnarray}
 To find the masses of the v-pions, only the two lightest
 flavours need to be considered when mass mixing can be
 neglected. For an $SU(2)$ flavour symmetry, one of the terms
 in eq~(\ref{hgo}) is redundant, since it can be transformed
 into one of the other terms using a determinant that is not dynamical:
\begin{eqnarray}
\mbox{Tr}[A^2]  &=&  \mbox{Tr}[A]^2 -  2\, \mbox{Det}[A]
\hspace{10mm} \mbox{($A$ is $2\times 2$ matrix)}
\end{eqnarray}

\noindent
Using this we may absorb the $c_3^{}$ term into $c_1^{}$. 
With these considerations, one obtains 
the v-pion masses  to $O(M^2)$ below:
\vspace{2mm}
\begin{eqnarray}
m_{\pi_v^\pm}^2  &\approx&  
m_{vp}^2
+ ~ (c_1^{} + c_2^{}) ~ \frac{m_a^4 + m_b^4
 + 4 \, m_a^{}  m_b^{} \, (m_a^2 + m_b^2)
 \cos \bar{\theta}_{\mbox{\tiny HV}}^{} + m_a^2 \, m_b^2 \,
 [2+4 \cos (2\bar{\theta}_{\mbox{\tiny HV}}^{})]}{m_a^2 + m_b^2 
+ 2 \, m_a^{} m_b^{} \cos \bar{\theta}_{\mbox{\tiny HV}}^{}} 
\nonumber \\
\nonumber \\
m_{\pi_v^0}^2  &\approx& 
m_{\pi^\pm}^2 + \frac{ (c_1^{} - c_2^{}) ~
 (m_a^2 - m_b^2)^2}{m_a^2 + m_b^2 + 2 m_a^{} m_b^{} 
\cos \bar{\theta}_{\mbox{\tiny HV}}^{}} 
\label{m2mass}
\end{eqnarray}

\medskip\noindent
which are correct up to
$\hspace{0mm} O \left[ c_1^{} 
\sin \bar{\theta}_{\mbox{\tiny HV}}^{}  \left( \frac{ m_a^{}
     m_b^{}}{m_{\pi_v^0}^{}} \right) \left( \frac{m_a^{} -
     m_b^{}}{m_a^{} + m_b^{}} \right) \right]^2 \vspace{1mm}$ terms.
For a discussion on the contributions from instanton effects 
to pseudo-Goldstone boson masses at $O(M^{N^\prime-1}) $, see 
\cite{JEKim} and references therein. This however does not affect 
the mass splitting of the v-pions.

\medskip
The best fit values for the coefficients of the higher order terms in
 QCD \cite{Pich:1995bw} determines $c_i^{\mbox{\tiny  QCD}}
 = 10^{-3} \, (\mu^2/f_\pi^2)^{\mbox{\tiny QCD}} \times [-0.6 \pm 0.3, 0.2 \pm 0.3, 
(0.9\pm 0.3])(N_v^{}/3)]_i^{}$, ($N_v=3$ for QCD),
where the $N_v^{}$ scaling for $c_i^{}$ 
follows in the large $N_v^{}$ limit \cite{Manohar}. In our case,  if 
the hidden sector physics can be approximated by a scaled up version 
of QCD, using eq (\ref{mvp}) and (\ref{m2mass}) with
$\bar{\theta}_{\mbox{\tiny HV}}^{} \sim 0$, a $10\%$ mass splitting
for the valley-pions would be found if
$(m_{a,b}^{} / \mu)^{\mbox{\tiny HV}} \gtrsim O(0.1)$. This can be
satisfied with a suitable choice of $m_{a,b}^{} ~(<
\Lambda_{\mbox{\tiny HV}}^{})$.
If $ \bar{\theta}_{\mbox{\tiny HV}}^{} \sim \pi$, the $\pi^\pm_v$
fields become massless in the limit of degenerate quark current
masses. However, the $\pi^0_v$ field remains massive, so it can
be very easy to find a $10\%$ mass splitting in this regime.
As  mentioned above, this mass splitting plays an important
 role in the discussion of the dark matter experimental constraints 
 on  the model, and we return to these in  Section~\ref{dmc}.

\section{Collider constraints on the $Z^\prime$ boson}\label{CZ'}

 In the following we examine the experimental constraints on 
 the parameters  describing the $U(1)^\prime$ extension of
 the SM presented in Section~\ref{HVM}, and investigate in particular
 the bounds on the mass of $Z'$. To this purpose we consider precision
  measurements at colliders and direct searches for new neutral gauge bosons.
 For a $Z^\prime$ mass of $O$(100)~GeV or less, additional constraints
 are relevant from determination of the muon anomalous magnetic moment
 \cite{Murakami:2001cs}. However, for the model considered this
 does not further restrict the limits presented.

\subsection{Electroweak precision data}\label{EWPD}

 In the following we use a model independent formalism
 \cite{Kennedy:1988sn}, 
 developed to quantify the deviation of physics from the Standard
 Model predictions. The parameters $\rho, x$ and $y$  introduced
 below define the low energy effective Lagrangian with the physical
 $W^\pm$ and $Z$ integrated out:
\begin{eqnarray}
{\cal{L}}_{\mbox{\tiny eff}} &=& - \frac{4 G_F^{} }{\sqrt2 \, g_W^2  
\sec^2 \theta_W^{}} 
\left[ \sec^2 \theta_W^{} \, J^{}_{W^+} \cdotp J^{}_{W^-} +  \rho \, J_{z}^2 
+ 2 x \, J_{z}^{} \cdotp J^\prime   + y \, J^{\prime \, 2}  
\right] ~+ \, \cdots
\label{lowe} 
\end{eqnarray}
where $J_{z}^{} ~=~ J_{W_3^{} }^{} - \sin^2 \theta_*^{} \,
J_{EM}^{}$. The modified Weinberg angle, $\theta_*^{}$, is introduced
to account for the extra component of the electromagnetic current that
mixing introduces. The experimental determination of the $\rho$
parameter assumes that the non-oblique terms 
 ($x,y$) are negligible. For the model discussed in Section~\ref{HVM}, 
the non-oblique coefficients $x,y$ presented below 
are determined at tree level, using eq~(\ref{jtrans}):
\begin{eqnarray}
x &=&  \frac{ \rho \, \sin \zeta \, \sec \chi}{ 
 \left( \cos \zeta + \sin \theta_w \tan \chi \sin \zeta \right) } 
\label{xdef}
\\ \hspace{10mm}
\nonumber \\
y &=& \frac{x^2}{\rho}
\label{ydef}
\end{eqnarray}
\vspace{0mm}
As $|\sin \chi | \rightarrow 1$ (maximal kinetic mixing),
 $|x| \rightarrow \rho \csc \theta_w$. 
In this limit, the product of the $U(1)^\prime$ coupling
and $U(1)^\prime$  charges of SM matter must be very small
in order to satisfy the experimental constraints.
The tree-level $\rho$ parameter in this model is:
\vspace{1mm}
\begin{eqnarray}
\rho &=&
\frac{m_W^{2} \, \sec^2 \theta_w }{m_{Z}^{2}  }  
 \, \left( \cos \zeta + \sin \theta_w \tan \chi \sin \zeta \right)^2
\nonumber \\[7pt]
&\approx& 1 +   \frac{m^{2}_W}{m^{2}_{Z^\prime}} 
  \frac{\tan^2 \theta_w}{\cos^2 \chi} 
\left[    \left(  2\ \! Q_H^\prime \, \frac{g^\prime}{g^{}_Y}      \right)^2
- \sin^2 \chi
\right]   
+O \left[   \frac{m^{4}_W}{ m^{4}_{Z^\prime}}  \right] .
\label{sinsym}
\end{eqnarray}
Once radiative effects in the SM are accounted for, the
 $\rho^{}_0 ~(\equiv \rho/\rho^{}_{SM})$ parameter, which is
 sensitive to physics beyond the SM, has been determined by a
 global fit to the EWPD giving $\rho_0^{} = 1.0002^{+0.0007}_{-0.0004}$ with
 the 1$\sigma$ limits shown \cite{pdg}.

\begin{figure*}[t]
\begin{minipage}{15.6cm}
\subfloat[$\alpha^\prime \equiv g^{\prime 2}/(4\pi)= 10^{-3}$]
{\includegraphics[width=6cm] {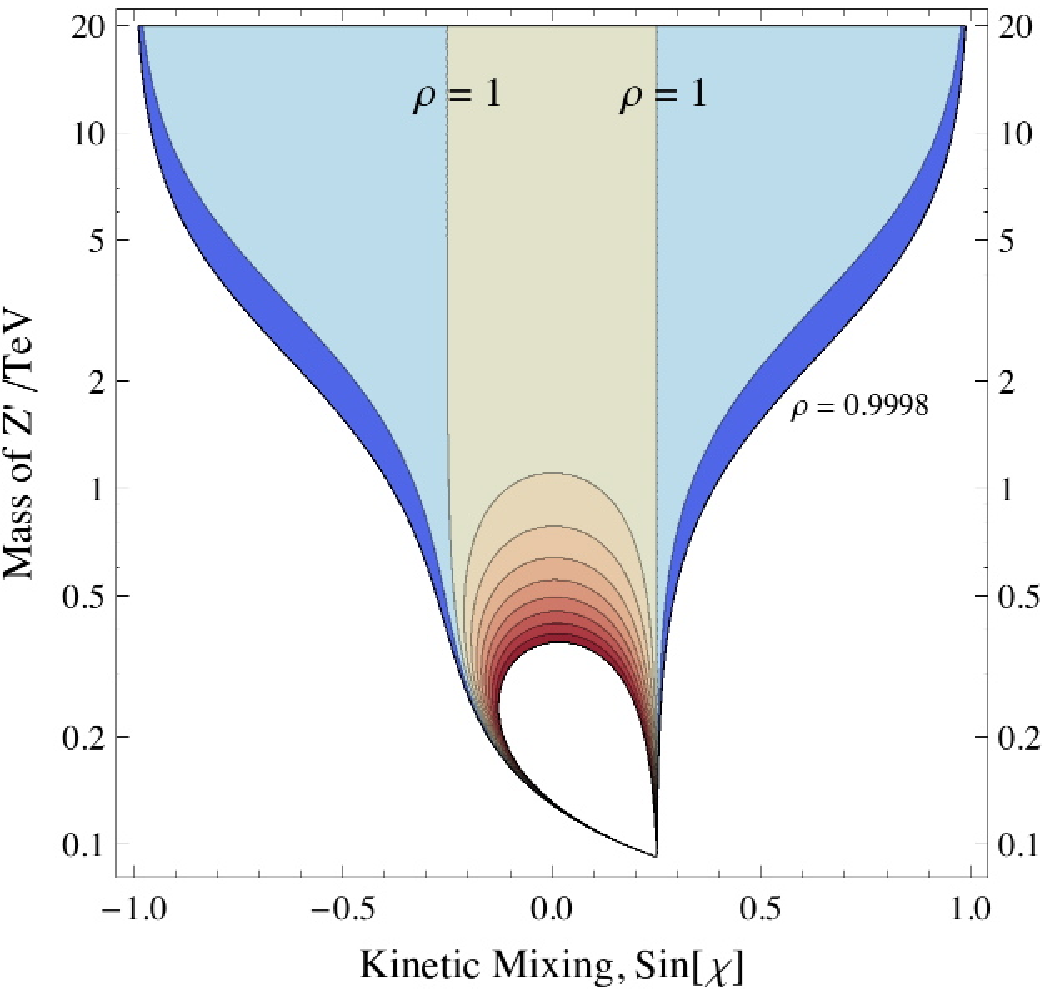} }
\hspace{0.4cm}
\subfloat[$\alpha^\prime\equiv g^{\prime 2}/(4\pi) = 1$]
{\includegraphics[width=5.9cm] {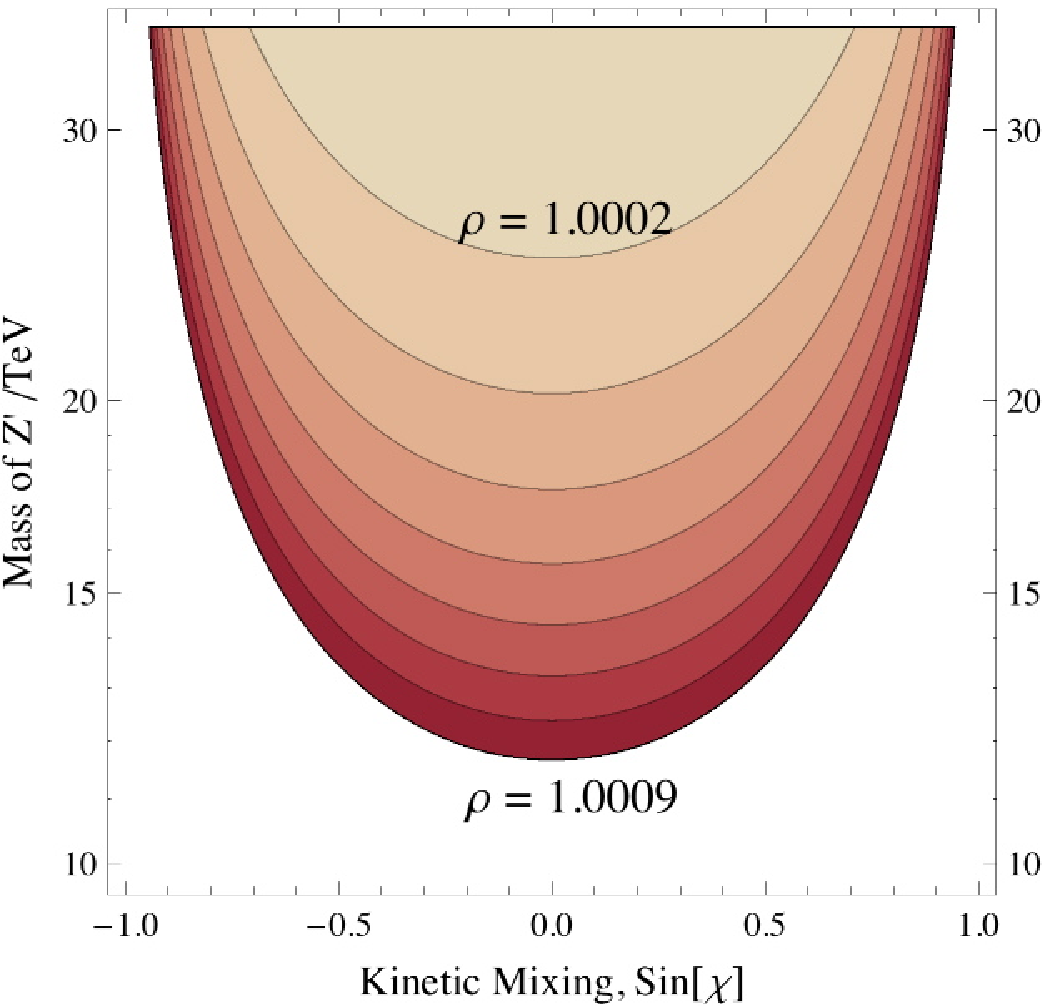} }
\hspace{0.4cm}
\subfloat{\includegraphics[width=2cm] {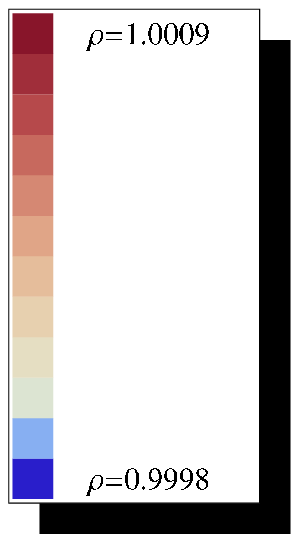} }
\end{minipage}
\def\baselinestretch{1.}
\caption{
Allowed parameter space for the $Z^\prime$ model 
given the measurement of the $\rho$ 
parameter from electroweak precision data, with $Q^\prime_H = 2/5$ 
(the contours of $\rho$ are spaced every $10^{-4}$).}
\label{alpS}
\end{figure*}

 Fig~\ref{alpS} shows
 the allowed parameter space given this constraint (with $Q_H^\prime
 =2/5$) for two values of $\alpha^\prime \,\equiv g^{\prime 2}/4\pi$.
From that one can easily extract the mass bounds on $Z'$ once we know the
values of the kinetic mixing, $\rho$ parameter and $\alpha'$.
Given the definition of the $\rho$ parameter 
in eq~(\ref{lowe}), one sees that the
 limits presented are insensitive to the hidden valley at tree level
 and also to the SM fermion $U(1)^\prime$ charges. In Fig~1a, the point
 where the current coupling to the physical Z is identical to the
 SM case $(\zeta = 0)$ is at $\sin \chi \sim 0.25$. The plot
 demonstrates that any value for $m_{Z^\prime}^{}$ will satisfy the
 $\rho$ constraint at this point when the tree level result is
 considered.
 For small $m_{Z^\prime}^{}$, although not noticeable in Fig 1a, the
 `arm' of allowed parameter space extending from the opposite sign of
 $\sin \chi$ does not meet with the straight $\rho=1$ line (ie there
 is no `hole' in the allowed parameter space). For large
 $m_{Z^\prime}^{}$,
 there is an approximate symmetry in $\chi \to - \chi$ of the allowed
 parameter space as suggested by eq~(\ref{sinsym}).

As $g^\prime \, Q_H^\prime$ is increased, the position of the straight
 $\rho =1$ line shifts to larger values for the kinetic mixing. The
 `arm' of the allowed parameter space also retracts, and the contours
 of $\rho$ shift to larger values of $m_{Z^\prime}^{}$. As
 $g^\prime Q_H^{\prime}$ is increased beyond $g_Y^{} /2$, the picture
 of the allowed parameter space then resembles Fig 1b and the contours
 continue to be shifted to larger values of $m_{Z^\prime}^{}$. It
 should be noted that when the $\rho$ constraint is very weak, other
 EWPD measurements provide stronger constraints. However,
 the strongest constraint on the model introduced in Section~\ref{HVM}
 is most often from either
 the $\rho$ parameter or direct searches.

\subsection{Direct $Z^\prime$ searches}\label{DZ'}

\begin{figure}[th]
\begin{tabular}{cc|cr|}
\parbox{7.cm}{\psfig{figure=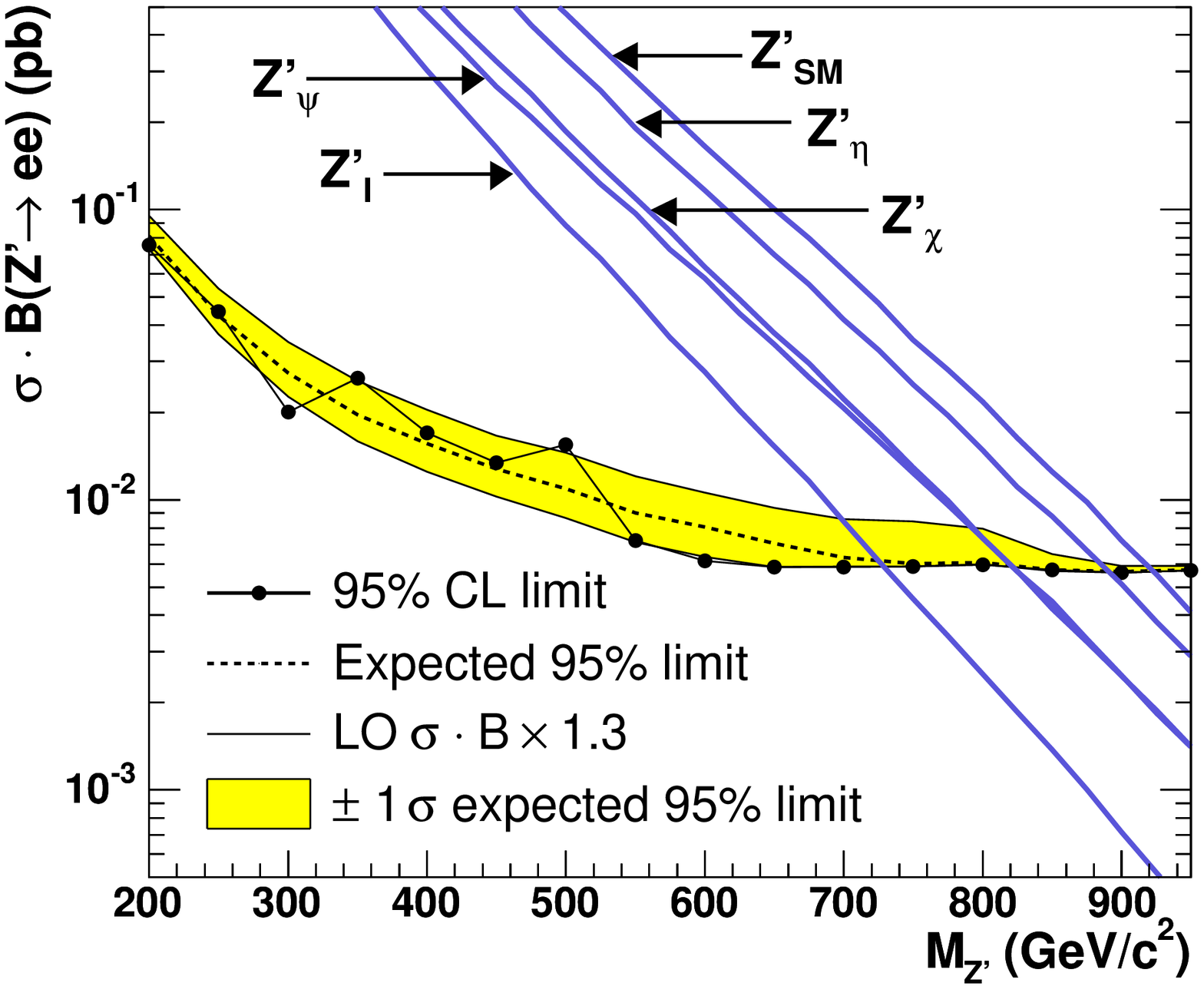,
    height=5.cm,width=6.7cm}
\def\baselinestretch{1.}
\vspace{0.1cm}
\caption{CDF limits on a new spin-1 \,\,\,
 particle (FERMILAB-PUB-07-367-E)}
 \label{cdf}}
\hspace{0.8cm}
\parbox{7cm}{
\vspace{0.45cm}
\psfig{figure=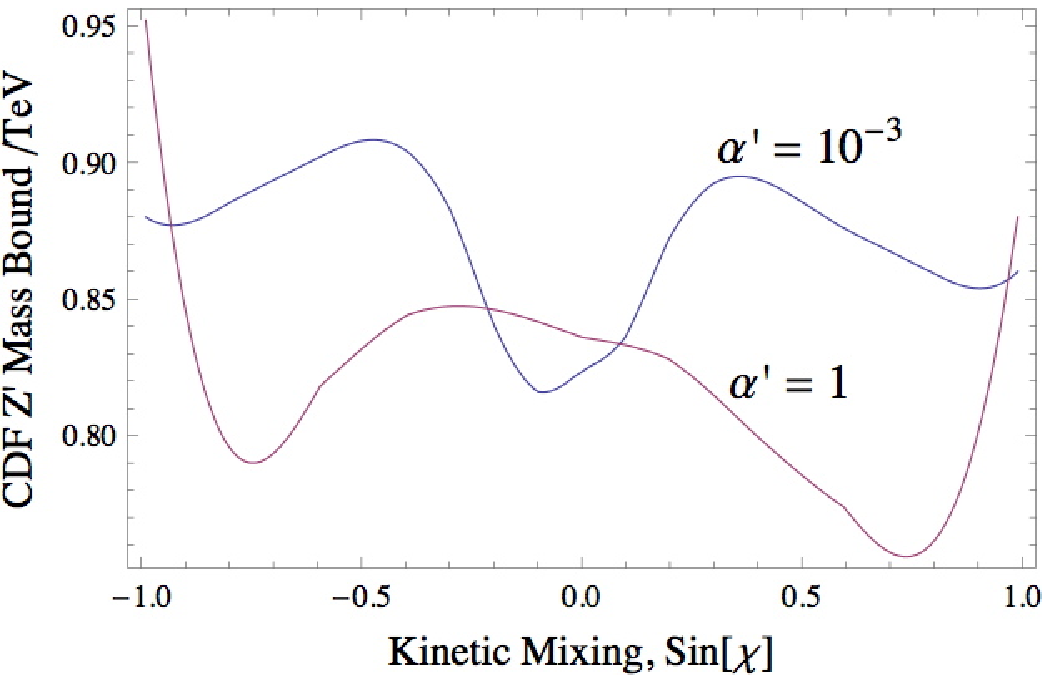, height=4.8cm,width=6.7cm}
\def\baselinestretch{0.99}
\vspace{-0.15cm}
 \caption{CDF lower mass bound on the $Z^\prime$ in the model 
presented in Section~\ref{HVM}.}
 \label{cdfb}}
\end{tabular}
\end{figure}

 In this section we consider the mass bounds produced by the Tevatron
 from searches for new gauge bosons beyond the SM. These can provide a
 more stringent lower bound for $m_{Z^\prime}^{}$ than electroweak
 precision data in some cases, and is still present when there is no
 $Z^\prime - Z^{}_{\mbox{\tiny SM}}$ mass mixing $(\zeta = 0)$. Fig
 \ref{cdf} shows their experimental limits
 \cite{cdfp}, and the theoretical cross-section for
 $p \bar{p} \rightarrow Z^\prime \rightarrow e^+ e^-$.

\noindent
 The boson labelled $Z^\prime_\chi$ has a lower mass bound of 822 GeV.
 Note that this  $Z_\chi'$  is the same as  $Z^\prime$ introduced in
 Section~\ref{U(1)}, but with no kinetic mixing and with the $U(1)^\prime$
 coupling fixed \cite{zth} to
 $g^\prime =  \sqrt{\frac{25}{24} } \, g_W^{} \tan \theta_w^{} $. It
 is also assumed that the branching fraction to invisible states is
 negligible. 
 Relaxing this assumption would reduce the theoretical cross section
 for producing SM final states, and so reduce this lower bound on the
 $Z^\prime$ mass. However, it will follow that this is a reasonable
 assumption for the hidden valley extension considered
 if the dark matter candidate discussed in Section~\ref{dmc}
 is the lightest HV state coupling to the $Z^\prime$, and the SM
 particles have similar (or greater) order of magnitude $U(1)^\prime$
 charges compared to the HV matter. The constraints on a $Z^\prime$
 where the SM particles have no $U(1)^\prime$ charges have been
 discussed in other
 papers \cite{Feldman:2006ce,Feldman:2006wb,Feldman:2006wd,
Feldman:2007wj,Feldman:2007nf}.

 The lower bound on the mass of the $Z^\prime$ introduced in
 Section~\ref{U(1)} for arbitrary kinetic mixing and negligible
 decay to HV states is shown in Fig~\ref{cdfb}. Since the
 theoretical resonant cross section is controlled by branching
 fractions, the lower bound on $m_{Z^\prime}^{}$ only varies
 by $\sim 10\%$ as the ($g^\prime, \chi$) parameters vary.
 For large $\alpha^\prime$, the $\rho$ constraint is much stronger.

\section{Dark matter constraints}\label{dmc}

Let us now address the dark matter constraints on the model.
 We first state the method for calculating the thermal relic
 abundance of a stable particle and discuss what conditions are
 necessary for a dark matter candidate from the ``hidden valley''
 to contribute significantly to the present day non-baryonic
 matter energy density. We present limits on the parameters of
 the hidden valley model
 from ensuring that dark matter is not  overproduced and consider
 the constraints from direct dark matter searches. We find in
 particular upper mass bounds on $Z'$ which will be combined with the
 lower mass bounds of Section~\ref{CZ'}.

\subsection{Thermal relic energy density}\label{RED}

During inflation, the reheating may have been sufficient for the
 $Z^\prime$ mediator to thermalise the SM and HV sectors. If not, the
 temperatures of the two sectors will be independent, and depend on
 the relative couplings to the inflaton. We consider the former case
 with a dark matter candidate from the hidden valley that
 annihilates into the SM sector.

The thermal relic density of a stable particle is controlled by when
 this particle leaves thermal equilibrium with its annihilation
 products (``freeze-out''). This occurs roughly when the thermally
 averaged annihilation rate of this particle becomes slower than the
 expansion rate of the universe:
\begin{equation}
n \, \langle \sigma_{\mbox{\tiny ann}} \, v \rangle \sim H
\label{fro}
\end{equation}
 where $n$ is the number density of the particle freezing out, $\langle
 \sigma_{\mbox{\tiny ann}} \, v \rangle$ is the thermally averaged
 annihilation cross section weighted by the relative velocity, and $H$
 is the Hubble constant. The ensemble of frozen-out particles then
 expand isenthropically, allowing the present number of particles in a
 comoving volume to be calculated. The current contribution to the
 energy density of the universe can then also be determined. If the
 particles are
 non-relativistic at the time of freeze-out, the thermal average is given by:
\begin{eqnarray}
\langle \sigma_{\mbox{\tiny ann}} \, v \rangle &\approx& 
\frac{x^{3/2} }{2 \sqrt\pi} \int_0^\infty dv \, v^2 \,
 (\sigma_{\mbox{\tiny ann}} \, v) \, e^{-x v^2/4} 
\label{therm}
\end{eqnarray}
where $x \equiv m_{\mbox{\tiny DM}}^{} / T$. The annihilation cross
 section can be calculated perturbatively, however for slow moving
 dark matter there can be important non-perturbative effects due to
 interactions before annihilation. This is accounted for by the
 Sommerfeld factor as discussed in detail in Appendix~\ref{appendixC}
 and other papers
 \cite{Sommerfeld,Hisano:2003ec,Hisano:2004ds,Hisano:2006nn,
Cirelli:2007xd,j1,MarchRussell:2008tu}.
 The temperature at which the freeze-out occurs is determined from
 eq~(\ref{fro}), and is parametrised by the value of $x$ at
 freeze-out \cite{Drees:2007kk}:
\begin{eqnarray}
x_F^{}   \approx \ln \left( \frac{\left[ 3.85 \times 10^{17} \,
 \mbox{GeV} \right] g_{\mbox{\tiny DM}}^{} \, m_{\mbox{\tiny DM}}^{}
 \langle \sigma_{\mbox{\tiny ann}} \, v
 \rangle_{x_F^{} } }{\sqrt{x_F^{} \, g_{*s}} } \right)
\label{xfreeze}
\end{eqnarray}
where $g_{\mbox{\tiny DM}}^{}$ is the number of degrees of freedom of
 the dark matter candidate. The $g_{*s}$ parameter is the effective
 number of relativistic degrees of freedom contributing to the entropy
 density of the universe. For $T \gtrsim 300$ GeV, $g_{*s} = 106.75$
 (assuming only Standard Model particles contributing), and at $T = 1$
 GeV, $g_{*s} \sim 80$. The contribution to
 the energy density is then given by:
\begin{eqnarray}
\Omega_{\mbox{\tiny DM}}^{} h^2 &=& \frac{8.6 \times 10^{-11}
 \,\, \mbox{GeV}^{-2} }{\sqrt{g_{*s} (x_F)} \, J(x_F) }
\hspace{7mm}
\mbox{where} \hspace{4mm}
J(x_F) ~=~ \int_{x_F^{}}^\infty dx \, x^{-2} \langle
 \sigma_{\mbox{\tiny ann}} \, v \rangle
\label{omh}
\end{eqnarray}
 where the approximation that all annihilations cease after
 ``freeze-out" has not been applied. The ``annihilation integral,"
 $J(x_F^{})$, accounts for the reduction in particle number after
 ``freeze-out." The thermal relic will contribute to the non-baryonic
 matter energy density observed in the universe today. The 5-year data
 from WMAP \cite{wmap1} suggests that $\Omega_{c}^{} h^2 = 0.1099 \pm
 0.0062$, so the central value will be used as an upper limit for the
 dark matter contribution to
 $\Omega \ \! h^2$.

 The thermal relic abundance increases as the annihilation cross
 section is reduced. Thus, whereas the constraints in
 Section~\ref{CZ'} led to small cross sections being preferred for
 $Z^\prime$ mediated or mixing events, the WMAP constraint will limit
 how small these cross sections
 are allowed to be, given a thermal relic from the hidden valley.

\subsection{Hidden valley dark matter candidate}\label{DMC}

 For the model discussed in Section~\ref{HVM}, there are no flavour
 changing interactions for the HV fermions. A state with a non-zero
 ``flavour" quantum number is therefore stable. If there are confining
 interactions for the HV matter, some of the composite states will be
 flavour neutral (eg $\pi_v^0$) and thus unstable due to the presence
 of the HV-SM mediator. However, there will also be stable
 states present (eg $\pi_v^\pm$) which are dark matter candidates. The
 lightest stable states dominate the relic density.
 Since  the last annihilation mechanism to freeze-out
 controls the relic density,  the freeze-out temperatures for
 the processes below must be evaluated and compared, to
 determine what the dominant annihilation mechanism is:

\begin{itemize}
\item
$~\pi^+_v \pi^-_v \to ~Z^\prime \, \to \, $ SM
\vspace{2mm}
\newline
 This process freezes-out at a temperature $T_1^{} \sim
 m_{\pi^\pm_v} /25$ when dark matter is not overproduced but the 
annihilation is not close to the $Z^\prime$ resonance (in which 
case $T_1^{} \sim m_{\pi^\pm_v} /40$). This was found using the 
annihilation cross sections to SM products listed in Appendix~\ref{appendixB}. 
The analysis in the following section
 assumes that this is the dominant process at freeze-out. For
 $\pi^\pm_v$ masses in the range $400$ GeV to $40$ TeV, the freeze-out
 temperature is passed at a time between $10^{-13}$ and  $10^{-8} \,
 \sec$. The $Z^\prime$ resonance is at $4 m_{\pi_v^\pm}^2 \sim
 m_{Z^\prime}^{} \sqrt{m_{Z^\prime}^2
 + \Gamma_{Z^\prime}^2} $ and this relates the upper mass bounds 
of the $\pi_v^\pm$ and $Z^\prime$ presented in Section~\ref{paramspace}.
\item
$\pi^+_v \pi^-_v \to ~\pi^0_v \pi^0_v  
$ ~(followed by decay to SM products)
\vspace{2mm}
\newline
 This annihilation proceeds via the strong interactions of the hidden
 sector, and may freeze-out after the $Z^\prime$ mediated
 annihilation. 
 Using the
 effective Lagrangian in eq~(\ref{lcpt}), the annihilation
 cross section is:
\begin{eqnarray}
\sigma^{}_{pp} &\sim&
 \frac{36  \pi^3}{\Lambda^2_{\mbox{\tiny HV}}}
 \left( \frac{m^2_{\pi^\pm_v}}{\Lambda^2_{\mbox{\tiny HV}}} \right)
 \frac{\beta_f^{} }{\beta_i^{} }
\end{eqnarray}
where the velocities of the initial and final states,
 $\beta_{i,f}^{}$, account for the phase space. A Sommerfeld
 enhancement from v-strong or other interactions could significantly
 raise this cross section further. The freeze-out temperature can
 be found using eq~(\ref{xfreeze}):
\begin{eqnarray}
T_2^{} &\approx& 
m^{}_{\pi^\pm_v}\, \Big/ \left[ 25 
- \ln \left( \frac{\Lambda^{}_{\mbox{\tiny HV}} }{10^6  \, \mbox{TeV}
} \right) 
- 3 \, \ln \left( \frac{\Lambda^{}_{\mbox{\tiny HV}}
}{m^{}_{\pi^\pm_v} 
  } \right) 
+ \ln \langle \beta_f^{} \rangle   \right]
\label{ttwo}
\end{eqnarray}
\vspace{-4mm}
\newline
The thermal average of the $\pi^0_v$ velocity is found by integrating
 over the kinematically
 allowed phase space, which in the non-relativistic limit gives:
\begin{eqnarray}
\langle \beta_f^{} \rangle &\approx& \sqrt{\frac{x}{\pi}} ~\left(
 \frac{  1 - \delta^2 }{ \delta } \right) ~e^{-\left(1-\delta^{2}
 \right) \, x /2} ~K_1^{}
 \left[ \left(1- \delta^{2} \right) x /2 \right]
\hspace{3mm} \mbox{if} \, ~\delta < 1
\end{eqnarray}
where $\delta = m_i^{} / m_f^{}$, and $K_1^{}$ is a modified Bessel
 function of the second kind. 
 If the SM sector does leave thermal equilibrium with the HV before the
 v-strong interactions freeze-out, the total energy in the hidden
 sector is reduced by the decay of the flavour neutral v-pion to SM
 states \cite{s1}. The relic abundance of the $\pi^\pm_v$ is then
 exponentially suppressed by the ratio of the time of freeze-out and
 $\pi^0_v$ lifetime, which can lead to a negligible
 contribution to the non-baryonic matter energy density. The $\pi_v^0$
 lifetime is given by:
\begin{eqnarray}
\Gamma_{\pi_v^0}^{-1} &\lesssim&  \Bigg(  \frac{10 \, \mbox{TeV}  }{
 \Lambda^{}_{\mbox{\tiny HV}}  }  \Bigg)^2 \Bigg( \frac{1 \,
 \mbox{TeV}}{m_{\pi_v^0 }^{} } \Bigg)
 \Bigg( \frac{m_{Z^\prime}^{}  }{ 10\, \mbox{TeV}  } \Bigg)^4 
 \times \frac{ 10^{-20} }{\alpha^{\prime 2} }  \,\, \sec^{} 
\label{pionlife}
\end{eqnarray}

\item If the v-pions are nearly degenerate with or heavier than the
 $Z^\prime$, the ``valley" feature of the hidden sector is lost, but
 extra annihilation channels can become
 important for the $\pi^\pm_v$ dark matter candidate. This
 situation is beyond the scope of the paper.
\end{itemize}

The unitarity limit places an upper bound on masses for a thermal
 relic at $O(10^2)$ TeV \cite{Griest:1989wd}. At this limit, we see
 from eq~(\ref{ttwo}) that the v-strong interactions will
 freeze-out before the $Z^\prime$ mediated interactions if either
 $\Lambda^{}_{\mbox{\tiny HV}} \gtrsim 10^3$ TeV or there is phase space
 suppression due to the $\pi_v^0, \pi_v^\pm$ mass difference.
 If the $\pi^0_v$ is heavier than the $\pi^\pm_v$, this process
 is known as a ``forbidden" channel \cite{Griest:1990kh}. 

For TeV~scale dark matter, we see that $T_2^{} > T_1^{}$ if either 
 $\Lambda^{}_{\mbox{\tiny HV}} /m^{}_{\pi^\pm_v} \gtrsim 30$ 
or there is phase space suppression, using eq (\ref{ttwo}). 
In the case $\Lambda^{}_{\mbox{\tiny
 HV}} / m^{}_{\pi^\pm_v}  =$ 8~(14), the $Z^\prime$ mediated
 annihilation will freeze-out last if the unstable v-hadrons are at
 least 10\%~(5\%) heavier than the strongly interacting dark matter
 candidate. It has been shown in Section~\ref{HVM} that this kind of
 mass spectrum can easily be realised in the hidden sector and in the
 following we assume that $T_2>T_1$ is indeed the case and the dark
 matter is $\pi_v^\pm$.

\subsection{Allowed parameter space and combined mass bounds on $Z'$}
\label{paramspace}

The constraints on a spin~0 thermal relic which dominantly annihilates
 via the $Z^\prime$ are now considered. In this the annihilation cross
 sections of Appendix~\ref{appendixB} are used, corrected by the
 Sommerfeld factors detailed in Appendix~\ref{appendixC}.
 Figures~\ref{dm3} and \ref{alph3} show the
 allowed parameter space for the dark matter and $Z^\prime$
 respectively,  for
 certain values of $U(1)'$ coupling $\alpha^\prime $ and charge $Q$, 
 with the SM $U(1)^\prime$  charges as given in Table~1.
 These plots assume that either the dark
 matter mass or the $Z^\prime$ mass is a free parameter to scan over,
 with $m^{}_{\mbox{\tiny DM}} < m^{}_{Z^\prime}$. If either of these
 parameters are further constrained by experiment, the parameter space
 of the other quantity will also be further constrained.

\begin{figure*}[t] 
\vspace{0mm}
\begin{minipage}{15.6cm}
\subfloat[~$\alpha^\prime \equiv g^{' 2}/(4\pi) = 10^{-3}$]
{\includegraphics[width=7cm] {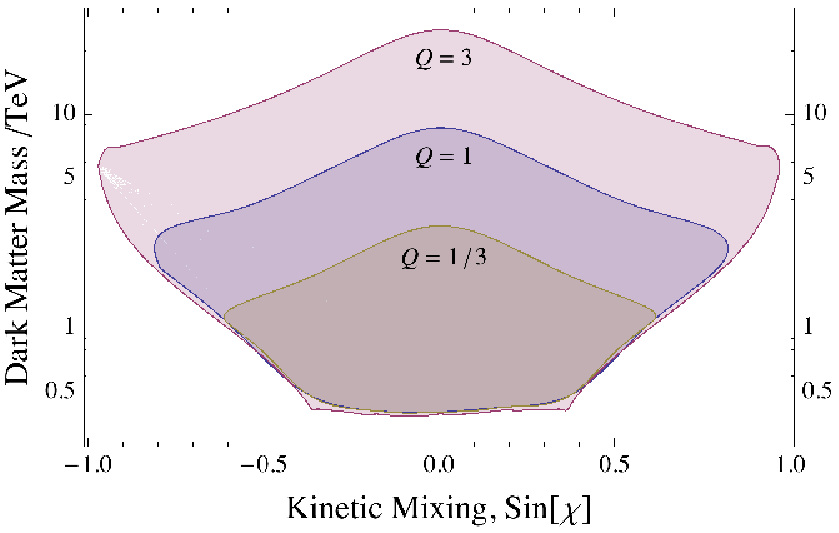} }
\hspace{10mm}
\subfloat[~$\alpha^\prime \equiv g^{' 2}/(4\pi) = 1$]
{\includegraphics[width=7cm] {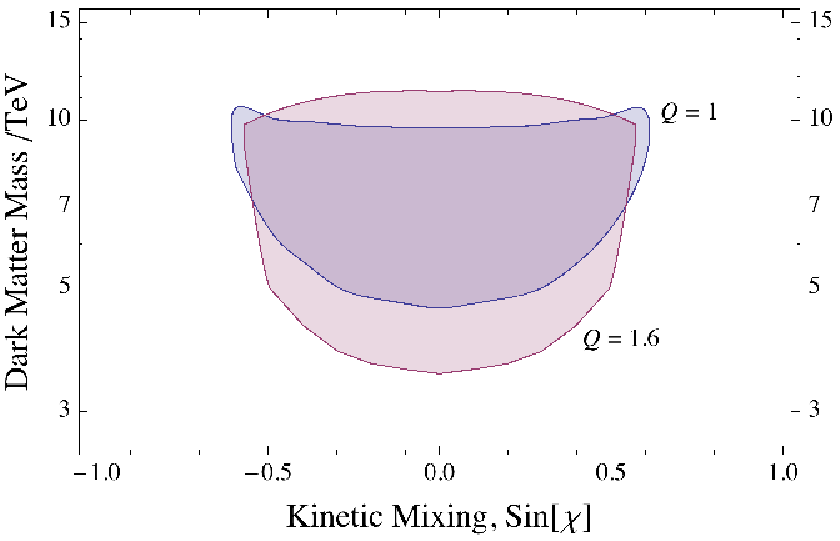} }
\end{minipage}
\def\baselinestretch{1.1}
\caption{Allowed parameter space for the spin 0
 dark matter with $U(1)^\prime$ charge, Q,
that dominantly annihilates via a heavier $Z^\prime$ with the SM
$U(1)^\prime$ charges given in Section~\ref{HVM}.}
\label{dm3}
\vspace{2mm}
\end{figure*}
\begin{figure*}[t]
\begin{minipage}{15.6cm}
\subfloat[~$\alpha^\prime \equiv g^{' 2}/(4\pi) = 10^{-3}$]
{\includegraphics[width=6cm] {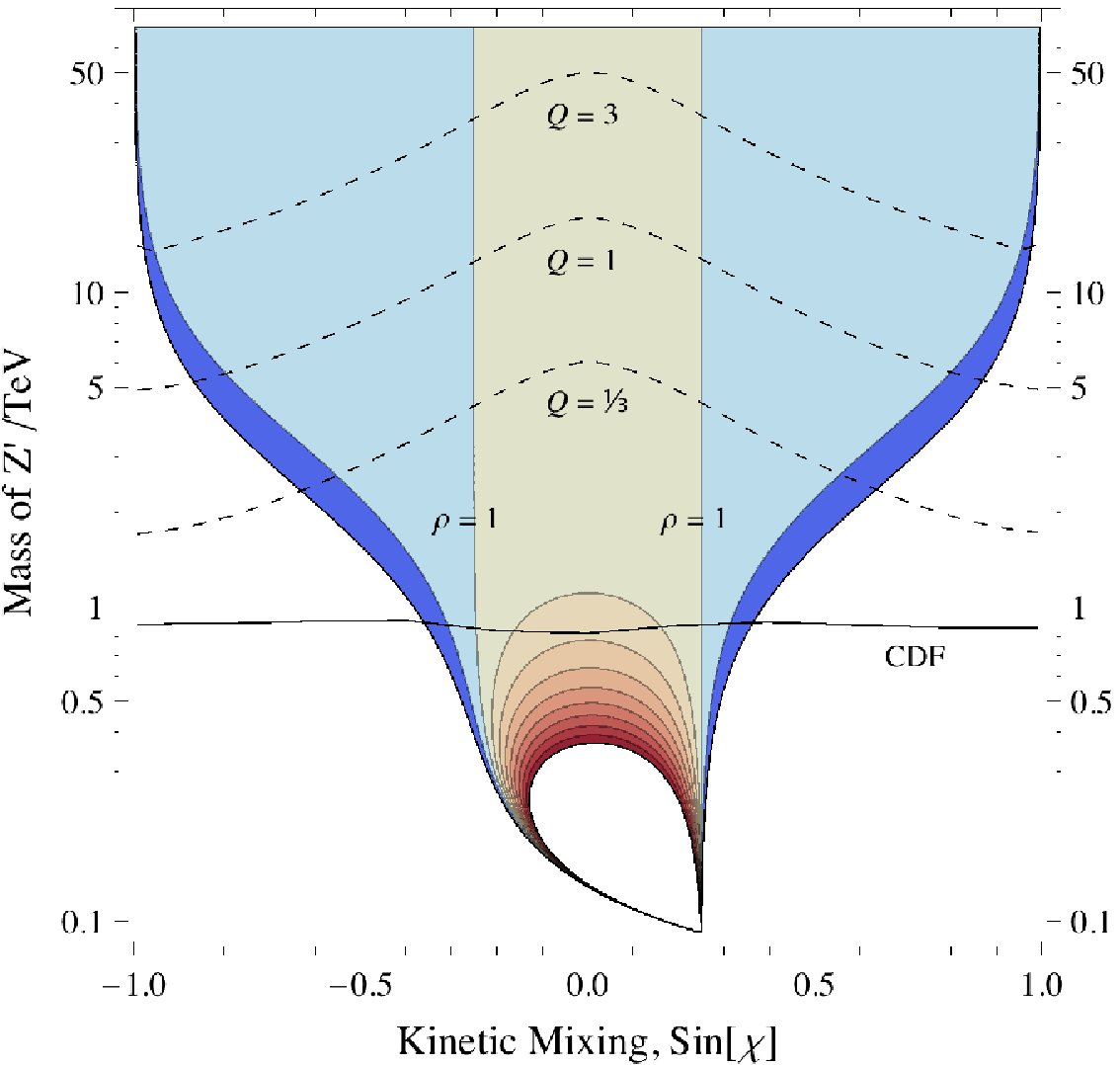} }
\hspace{10mm}
\subfloat[~$\alpha^\prime \equiv g^{' 2}/(4\pi) = 1$]
{\includegraphics[width=5.9cm] {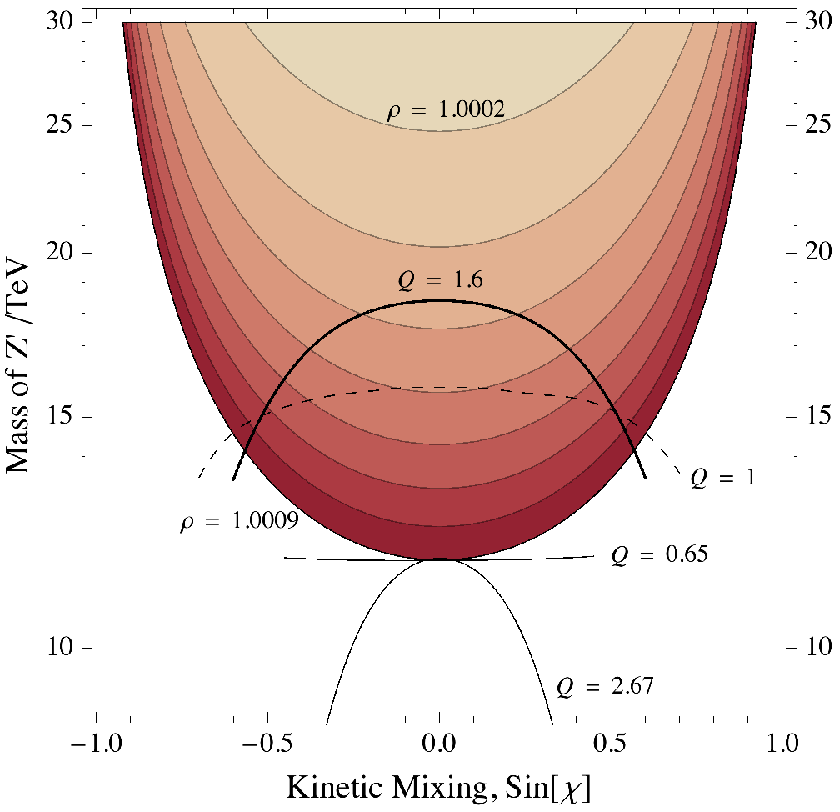} }
\hspace{0.8mm}
\subfloat{\includegraphics[width=1.8cm] {rlegend} }
\end{minipage}
\def\baselinestretch{1.1}
\caption{Combined constraints on the $Z^\prime$ with the coloured
 regions showing the $\rho$ parameter allowed space with $1 \sigma$
 limits determined from electroweak precision data, and lines for the
 lower mass bound from $Z^\prime$
 searches at CDF and upper mass bounds given the presence of a spin~0
 dark matter candidate with $U(1)^\prime$ charge, Q, that dominantly
 annihilates via a heavier $Z^\prime$.
(The contours of the $\rho$ parameter are spaced every $10^{-4}$).}
\label{alph3}
\vspace{2mm}
\end{figure*} 

The upper mass bound on the $Z^\prime$ is found when the dark matter
 annihilation cross section on the $Z^\prime$ resonance is below a
 critical point, leading to over-production of the DM candidate with
 respect to the WMAP constraint. The lower mass bound of the dark
 matter allowed parameter space is controlled by the lower bound of
 the $Z^\prime$ mass and is the analogue of Lee-Weinberg bound for 
heavy neutrinos \cite{Lee:1977ua}. If the CDF constraint on the 
$Z^\prime$ mass is
 relaxed due to a significant branching fraction into HV states, it
 can be possible for the dark matter
 to be lighter than that shown in Fig~\ref{dm3} for small kinetic
 mixing ($|\sin \chi| \lesssim 0.4$).

As $\alpha^\prime Q_{\mbox{\tiny eff}}^2$ increases, where
 $Q_{\mbox{\tiny eff}}^{}$ is the coupling of the $\pi^\pm_v$ to the
 $Z^\prime$ which accounts for kinetic and mass mixing using 
eq (\ref{jtrans}), the effect of interactions before annihilation
 (the Sommerfeld effect) becomes much more significant. For spin~0
 dark matter, the repulsive interaction between the dark matter
 particles leads to an exponential suppression of the annihilation
 cross section. So, the allowed parameter space cannot be made
 arbitrarily large by increasing the $U(1)^\prime$ charge of the dark
 matter. When $\alpha^\prime =1$ and the other parameters are
 unrestricted, the greatest value for  $m_{Z^\prime}^{}$ consistent
 with the WMAP determination of $\Omega_c^{} h^2$ is $\sim 18$ TeV. As
 the coupling of the dark matter to the physical $Z^\prime$ is
 proportional to $\sec \chi$, the Sommerfeld effect also produces much
 stronger limits on the amount of kinetic mixing allowed than expected
 from perturbative calculations.

 The upper limits on the masses of the spin~0 dark matter and
 $Z^\prime$ are found where the dark matter candidate saturates the
 non-baryonic matter energy density, $\Omega_c^{} h^2$. If a separate
 particle is found at the LHC, which contributes to $\Omega_c^{} h^2$,
 the allowed parameter space of the HV dark matter candidate and
 $Z^\prime$ will be reduced. However, if the annihilation cross section of
 the HV dark matter has been underestimated, the upper limits are
 raised. This could be the case from having neglected annihilation
 products involving supersymmetric SM states.

 It is also possible for the HV dark matter to dominantly annihilate
 through the physical Z, due to kinetic and mass mixing. This does not
 change the bounds on the physical $Z^\prime$ mass, but does sometimes
 allow a very narrow region of parameter space for the dark matter
 around the $Z$ resonance, $m_{\mbox{\tiny DM}}^{} \sim m_Z^{} /2$.
 This has not been shown in the plots.

\subsection{Direct dark matter searches}\label{ddms}

\begin{figure}[t]
\center
\includegraphics[width=11cm] {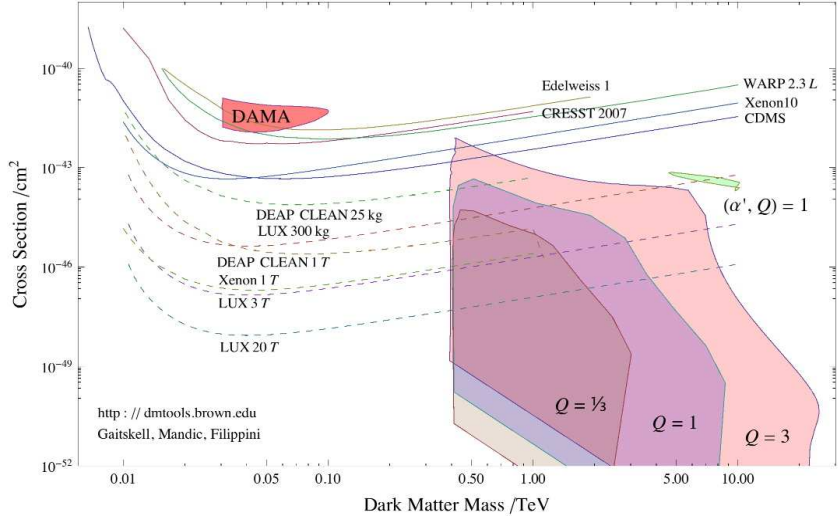}
\caption{Spin independent cross section for the dark
 matter scattering off a proton  
($\alpha^\prime = 10^{-3}$ apart from green region where $\alpha^\prime = 1$)}
\label{drct0}
\end{figure}

This section considers the constraints from direct dark matter
 searches on the HV dark matter candidate. There are numerous
 experiments which have looked for recoils of nuclear matter due to
 collisions with dark matter. There have not been any signals
 observed, except for an annual modulation in the DAMA experiment
 which prefers light dark matter. The dark matter in this model is too
 heavy to explain the DAMA signal, so the upper limits on the
 scattering cross sections from the other experiments are
 considered. The current and projected experimental limits are shown
 in Fig \ref{drct0} with solid and dashed lines respectively. These
 cross section limits assume a local dark matter density of $\sim 0.3$
 GeV cm$^{-3}$ and mean velocity of $\sim 220$ km s$^{-1}$. If the
 dark matter candidate does not saturate the
 non-baryonic matter energy density, then the 
 experimental limits are weakened.

Fig~\ref{drct0} also shows the theoretical prediction from scans of
 the allowed parameter space, as given in
 Section~\ref{paramspace}. For the allowed dark matter masses, the
 Sommerfeld effect is negligible for these calculations. This is
 because the $Z^\prime$ is much heavier than
 a proton so the interaction is screened to a small distance
 compared to the characteristic scattering length scale.

For a certain choice of parameters, the effective vector coupling of
 the proton to the $Z^\prime$ is zero. In the neighbourhood of this
 region, the dominant spin independent scattering occurs via the
 $Z$. Since the dark matter only gains milli-$Z$-charges from mixing,
 the cross section is suppressed. The vast majority of parameter space
 gives a theoretical cross section within a factor of 10 of the upper
 boundaries for the regions shaded in Fig~\ref{drct0}. It will be
 unlikely that all experiments use atoms for the dark matter to
 scatter off that are in the special region of parameter space that
 allow the very small cross sections.

 \subsection{Indirect dark matter searches}\label{idms}
 
In this section, we briefly review the prospects for indirect
detection of the HV dark matter candidate. There are searches for
signals of dark matter annihilation or decay in our galactic
neighbourhood. For the latter, the lifetime of the dark matter
candidate should be greater than the lifetime of the universe in order
for there to be a significant energy density currently present in the
universe. Some possible signals include highly energetic photons,
neutrinos and antimatter subject to a propagation dispersion
relation. The stability of the HV dark matter candidate was inferred
by the presence of a flavour symmetry. However, gravity is flavour
blind and so we expect  Planck suppressed operators to
break this symmetry and mediate decay of the $\pi_v^\pm$. As for dark
matter annihilation signals, since the average velocity of cold dark
matter in the galactic halo is of $O(10^{-3})$, the Sommerfeld
suppression for the v-pions to annihilate via the $Z^\prime$ strongly
reduces the cross section. This results in a negligible signal for
current experimental detectors. For dark matter that aggregates inside
stars, large planets or dwarf satellite galaxies where the velocities
are $O(10^{-5})$, the annihilation signal is even further
suppressed. Thus for the repulsive interaction considered here the
prospects for indirect dark matter searches are poor, in contrast to
the case considered in \cite{j1} that for an attractive
interaction between the dark matter candidates, these objects can
provide a much more favourable source for dark 
matter annihilation signals.

\section{Conclusions}
 
We considered experimental electroweak and dark matter constraints
 on a simple  extension of the SM with an additional $U(1)'$ 
massive gauge boson  and a particular type of a hidden sector (``hidden
valley'') with a confining, QCD-like gauge group and hidden valley matter 
charged under the new  $U(1)'$.
 As a result the SM and hidden valley sectors communicate via renormalisable
 operators involving  $Z^\prime$  as a messenger,  which can then probe the
 physics of the hidden valley sector. The latter can also provide a 
 dark matter candidate.  Combined electroweak and dark matter
 constraints placed both lower and upper bounds on the mass of $Z'$ 
 and these were studied in detail.

 We found from corrections to the $\rho$ parameter in electroweak
 precision data and direct searches a lower limit on the $Z^\prime$
 mediator mass of O(1-10 TeV) depending on the $U(1)^\prime$ coupling
 constant and kinetic mixing parameter. This result followed when
 there is a negligible $Z^\prime$ partial decay rate to HV states and
 $O(1)~U(1)^\prime$ charges for the SM matter. This limit does not
 require the presence of a hidden valley, as it would similarly apply
 for a pure $U(1)^\prime$ extension to the SM.  In addition, it was
 demonstrated that the kinetic mixing of the $U(1)^\prime$ and
 hypercharge gauge bosons is not restricted to be small by current
 observations. 

 We also found a dark matter candidate in the strongly interacting
 hidden valley that could saturate the present day non-baryonic matter
 energy density. If the spin~0 HV thermal relic dominantly annihilates
 via the $Z^\prime$, and the $U(1)^\prime$ charges of the SM and HV
 states are of similar magnitude, there is an upper bound on the
 $Z^\prime$ mass of O(10) TeV and restrictions on the amount of
 kinetic mixing allowed. Although the upper limit is beyond the reach of the
 LHC \cite{Feldman:2006wb,Cvetic:1995zs}, a large proportion
 of the parameter space will be tested. Detection of an
 alternative candidate to contribute to the dark matter density at the
 LHC could forbid the presence of this HV thermal relic. The lightest
 HV state was found to annihilate close to the $Z^\prime$
 resonance, restricting it to have an $O$(TeV) mass. Future direct
 dark matter searches will also significantly probe the currently
 allowed dark matter parameter space. 

 The inclusion of the Sommerfeld effect for p-wave annihilations of
 the spin~0 dark matter via a spin~1 state brought a significant
 change to the annihilation cross section. This constrained the
 parameter space much more than perturbative calculations would have
 suggested.

 If the lightest HV states are O(TeV) and the model discussed is
 realised, then extra signals containing highly energetic 
 SM decay products of HV states may be observed in
 collider detectors. The pseudoscalar $\pi_v^0$ state would
 preferentially decay to top quarks, gauge bosons and Higgs fields, and the
 LHC\,/\,Tevatron would see low multiplicity events. There would
 be insufficient energy and resolution to see the influence
 of the non-pseudo-Goldstone boson states. In alternative hidden valley
 scenarios, the phenomenology may be different
 and further study is important to identify how such
 models and their collider signals can be constrained.

\section*{Acknowledgements}

This work was partly supported by the EU
 contract MRTN-CT-2006-035863.
SC is supported by the UK Science and Technology Facilities Council 
(PPA/S/S/2006/04503). We thank John March-Russell, 
Subir Sarkar and Stephen West  for  interesting 
discussions and suggestions, Richard Gaitskell and 
Jeffrey Filippini for providing the aggregated  data 
sets of experimental limits from
 direct dark matter searches  and 
Jihn E. Kim for a clarifying discussion.

\section{Appendix}

\def\theequation{A-\arabic{equation}}
\def\thesubsection{A}
\setcounter{equation}{0}
\def\thefigure{A-\roman{figure}}
\setcounter{figure}{0}

\subsection{GUT embedding and neutrino masses}
\label{appendix0}

We provide here further details to  the  discussion in
Section~\ref{HVM} to show how 
neutrino masses can be generated in the model considered.
The model discussed may be considered a low energy realisation
of a theory with $SO(10)$ symmetry
if the SM Higgs field is part of a {\bf{10}}, {\bf{120}} or
 $\mathbf{\overline{126}}$ 
representation of $SO(10)$, and the non-SM states have quantised
 $U(1)^\prime$ charges, $q^{}_\pm , Q^\prime_\phi = 0, \pm1$, or
$\pm2$, also forming parts of $SO(10)$ multiplets. The missing
 states of these $SO(10)$ multiplets can have masses of the
 $SO(10)$ unification scale depending on the mechanism that breaks 
this symmetry.

To generate a large Majorana mass for the right-handed neutrino via 
the Higgs mechanism we choose $Q^\prime_\phi = 2$, and so if
unification is desired $\phi$ must belong to a
$\mathbf{\overline{126}}$ 
representation, which may be the multiplet $H$ is embedded in. The 
vacuum expectation value  of $\phi$ is also responsible for 
generating a mass for the $Z^\prime$ which we will find is $O($TeV$)$  
scale. The neutrino mass matrix for one SM family is of the form:
\medskip
\begin{eqnarray}
\left( \begin{array}{cc} \nu  &  N^c \end{array}\right)
\left( \begin{array}{cc} 0 & y^{}_\nu \, \langle H \rangle  \\  
y^{}_\nu \, \langle H \rangle & \lambda \,  \langle \phi \rangle
 \end{array}\right)
\left( \begin{array}{c} \nu  \\  N^c \end{array}\right)
\label{nmtx}
\end{eqnarray}
 
\begin{figure}[ht]
\center
\includegraphics[height=3.5cm,clip,trim = 0mm 1mm 0mm 0mm] {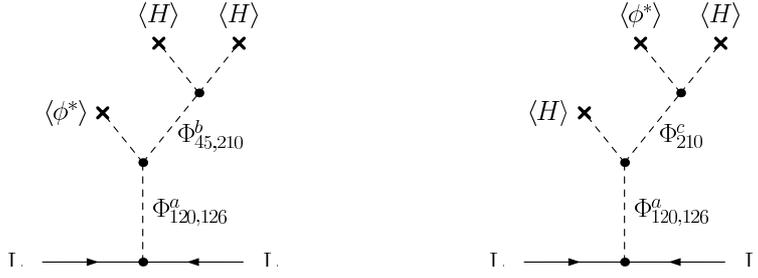}
\def\baselinestretch{1.}
\caption{Some Feynman diagrams that lead to an effective Majorana mass
 for the SM left-handed neutrinos after integrating out heavy scalar
 states, $\Phi$. The ``X,Y" label on $\Phi_{X,Y}^{}$ indicates that
 this scalar can be considered as part of an $\mathbf{X}$ or
 $\mathbf{Y}$ representation of SO(10). The $\Phi^a$ field has the
 representation $(\mathbf{1}, \mathbf{1})(1,-\frac65)$ under SU(3)
 $\times$ SU(2) $\times$ U(1)$_Y^{} \times$ U(1)$_\chi^{}$, $\Phi^b$
 is $(\mathbf{1}, \mathbf{1})(-1,-\frac45)$, and $\Phi^c$ is
 $(\mathbf{1}, \mathbf{2})(-\frac12,\frac85)$. }
\label{nmass}
\end{figure}

\medskip\noindent
The relation of the scalar vevs to gauge boson masses and couplings 
is given in Appendix~\ref{appendixB}. For $\lambda$ of $O(1)$, the 
heavier neutrino has a mass similar to or heavier than the $Z^\prime$ 
boson, and for a phenomenologically acceptable mass of $O(eV)$ for 
the light neutrino, the Yukawa coupling leading to the Dirac mass 
term must be small, $y_\nu^{} \sim 10^{-5}$. 
This is not much smaller than the electron Yukawa coupling. However, 
if one wants to avoid the introduction of such a coupling, 
a $Z_2^{}$ symmetry may be introduced as 
suggested in \cite{s1} with only the $N_i^{}$ fields having odd
parity (in this case the $N_i$ fields are dark matter candidates). 
This assignment is discrete anomaly free if heavier states with odd 
parity have been integrated out \cite{Ibanez:1991hv}, and it forbids 
the Dirac mass term. A Majorana mass could be generated for the 
left-handed neutrinos by a $G_{\mbox{\tiny SM}}^{} \times U(1^\prime)$ 
invariant dimension~6 operator, $LL HH \phi^* / M_\star^2$. If this 
operator is responsible for the light neutrino masses, it suggests 
that the scale of new physics which leads to lepton number violation 
is $M_\star^{} \sim 10^8$~GeV. Figure~\ref{nmass} shows some mechanisms 
for generating this dimension~6 operator.

\def\theequation{B-\arabic{equation}}
\def\thesubsection{B}
\setcounter{equation}{0}
\def\thefigure{B-\arabic{figure}}
\setcounter{figure}{0}
\subsection{Gauge boson eigenstates}\label{appendixA}

Here we provide additional details to the discussion in
Section~\ref{U(1)}, by computing
the gauge boson eigenstate fields in the presence of various  
mixings present.  The kinetic mixing in
\begin{eqnarray}
{\cal{L}}_{\mbox{\tiny gauge kinetic}}^{}  ~=~
-\frac14 \left( F_Y^{} \cdotp F_Y^{} + F_{}^{\prime} \cdotp
F_{}^{\prime}
 + 2 \sin \chi \, F_{}^{\prime} \cdotp   F_Y^{} \right) +
 \mbox{non-Abelian kinetic terms}.
\end{eqnarray}
is removed by the redefinition: 
\begin{eqnarray}
\left( \begin{array}{c} B_\mu  \\  B^{\prime}_\mu  \end{array}\right) =
\left( \begin{array}{cc} 1  &  - \tan \chi  \\ 0 &  \sec \chi 
\end{array}\right) 
\left( \begin{array}{c} \tilde{B}_\mu  \\  \tilde{B}^{\prime}_\mu  
\end{array}\right)
\label{redef}
\end{eqnarray}
where $B_\mu$ is the SM hypercharge  boson and $B^\prime$ is the 
$U(1)^\prime$ gauge boson. After this, $\tilde B_\mu$ and $\tilde B_\mu'$
have  canonical  kinetic terms. Under this transformation
 ($\mathbf{b} = \mathbf{C} \, \mathbf{\tilde{b} }$) the source terms
 for the gauge fields in the Lagrangian, $\mathbf{b}^T \, \mathbf{j}$,
 must be invariant, and this fixes the relation between new (``tilded'') 
and original currents: $\mathbf{\tilde{j} } = \mathbf{C}^T \,
 \mathbf{j}$. It then follows that with the field redefinition in
 eq~(\ref{redef}), the couplings to $\tilde B_\mu$ 
 (the new hypercharge) are the same as in the original basis.
After EWSB it is convenient to introduce the following fields:
\medskip
\begin{eqnarray}
\left( \begin{array}{c} \tilde{A}^{}_\mu  \\  \tilde{Z}^{}_\mu  \\ 
 \tilde{Z}^{\prime}_\mu  \end{array}\right) &=&
\left( \begin{array}{ccc} \cos \theta_w & \sin \theta_w & 0  \\ 
 -\sin \theta_w & \cos \theta_w & 0 \\ 0&0&1  \end{array}\right)
\left( \begin{array}{c} \tilde{B}^{}_\mu  \\  W^{3}_\mu \\
 \tilde{B}^{\prime}_\mu \end{array}\right) 
\end{eqnarray}

\noindent
The $\tilde{A}_\mu$ field will have no mass terms since it couples
to the generator $T_3^{} + Y$ by definition of the Weinberg angle
($U(1)_{EM}$ unbroken).
The SM Higgs field couples to both the SM neutral gauge boson
and to $B^\prime$, so after spontaneous symmetry breaking there 
will be mass mixing  terms. In general, mass mixing terms
still exist after going  to the ``tilde'' basis:
\medskip
\begin{eqnarray}
\frac{1}{2}
\left( \tilde{Z}^{}_\mu  \hspace{2mm}  \tilde{Z}^{\prime}_\mu \right)
\left( \begin{array}{cc} m_{}^2  &  -\Delta_{}^{}  \\
 -\Delta_{}^{} &  M_{}^2 \end{array}\right)
\left( \begin{array}{c} \tilde{Z}^{\mu}  \\  \tilde{Z}^{\prime \mu}  
\end{array}\right)
+ W^+ W^- \hspace{1mm} \mbox{term}.
\end{eqnarray}

\noindent
The rotation necessary to diagonalise this mass matrix is
 parametrised by $\zeta$ as follows:
\medskip
\begin{eqnarray}
\left( \begin{array}{c} Z_-^{}  \\  Z_+^{} \end{array}\right)
&=&
\left( \begin{array}{cc} \cos \zeta  &  -\sin \zeta  \\
 \sin \zeta &  \cos \zeta \end{array}\right)
\left( \begin{array}{c} \tilde{Z}^{\mu}  \\  \tilde{Z}^{\prime \mu}  
\end{array}\right)
\end{eqnarray}

\medskip\noindent
where $Z_\pm^{}$ are the mass and kinetic eigenstates, 
called the physical $Z$ and $Z^\prime$ in the main text.
With these relations one finds eq.(\ref{jtrans})
in the text for the transformation of the currents.

\def\theequation{C-\arabic{equation}}
\def\thesubsection{C}
\setcounter{equation}{0}
\def\thefigure{C-\arabic{figure}}
\setcounter{figure}{0}
\subsection{Cross section formulae}\label{appendixB}

We list here the partial decay rates and cross section formulae used
to generate the limits presented in the text, Section~\ref{dmc}.
 The interaction terms considered are parametrised below, with
 canonically normalised kinetic terms assumed:
\begin{eqnarray}
&  i g^{}_X Q^{}_X \, \phi \, (\partial_\mu \phi^*) \, X^\mu 
 + g_{X X^\prime }^{} \, v_\phi \,\, X^\mu X^{\prime}_{\mu } \,\,  \phi 
- g^{}_X \, \bar{\psi}  \, \gamma^{}_\mu X^{\mu} \left(  c^{}_V
 - c^{}_A \gamma^{}_5  \right)^X \psi  + \hc & \\
\nonumber \\
& \mbox{where} ~~g_{X X^\prime }^{} = 
\sqrt{2}  \,\, g_X^{} g_{X^\prime}^{} \, Q_{X  \, (\phi)}^{ } \,
 Q_{X^{\prime}  \, (\phi)}^{ } &
\end{eqnarray}
where $\phi$ is a spin~0 field, $v^{}_\phi = \sqrt2 \langle \phi \rangle$,
 $X^\mu, X^{\prime \mu}$ are gauge fields and $g_{X}^{},
 g_{X^\prime}^{}$ are the respective coupling constants. $Q_{X  \,
 (\phi)}^{}$ is the charge of the $\phi$ field under the symmetry
 whose gauge boson is $X^\mu$ and so on. The decay width of the
 $Z^\prime$ is an important quantity for any $Z^\prime$ mediated
 event close to the propagator pole. Some partial widths are given
 below, with the others assumed to be negligible. When relevant, the
 final state spin or polarisation configurations are summed over (but
 not for different
 colours).

\begin{eqnarray}
\Gamma_{X^{} \rightarrow f \bar{f}} ~~~~\,\,\, &=&
\frac{g_X^2 \,m_X^{}}{12 \pi } \sqrt{ 1 -  \frac{4 m_f^2}{m_X^2}  }
\, \left[ (c_V^2)^X \left( 1 +  \frac{2m_f^2}{m_X^2} \right)  +
  (c_A^2)^X 
\left( 1 -  \frac{4m_f^2}{m_X^2} \right)  \right]\qquad\qquad\quad \\
\nonumber \\
\Gamma_{X^{} \rightarrow \phi \phi^*} ~~~~&=&
\frac{ g_X^2  \, m_X^{} \, Q_{X  \, (\phi)}^{\, 2} }{48 \pi}
 \left( 1 -  \frac{4 m_\phi^2}{m_X^2} \right)^{3/2} \nonumber 
\nonumber
\end{eqnarray}
\begin{eqnarray}
\Gamma_{X \rightarrow W^+ W^-} &=&
\frac{g_X^2 \, m_X}{192 \, \pi} \left( 1-\frac{4m_W^2}{m_X^2}
\right)^{3/2} \! \left(  \frac{g_W^2 \, c_X^2}{g_X^2}  \right)
\left( \frac{m_X }{ m_W } \right)^4 \left[ 1\! +\! 20 \left( \frac{m_W }{
    m_X } \right)^2\! +12 \left( \frac{m_W }{ m_X }
 \right)^4 \right] \nonumber \\[10pt]
\Gamma_{X^{} \rightarrow {X^\prime} \phi} ~~~~ &=&
\frac{g_{X X^\prime }^2 \, m_X}{192 \, \pi  } \,
\left( \frac{v_\phi^2}{  m_{X^\prime}^2 } \right)
\sqrt{ 1-  2 \left( \frac{ m_{X^\prime}^{2} + m_{\phi}^{2} }{m_X^{2} }
 \right) +  \left( \frac{ m_{X^\prime}^{2} - m_{\phi}^{2} }{m_X^{2} }
 \right)^2 } \times \nonumber \\
&& \hspace{45mm} ~\left[  1 +2 \left( \frac{ 5 \, m_{X^\prime}^{2} -
    m_{\phi}^{2} }{m_X^{2} } \right)  + \left( \frac{ m_{X^\prime}^{2}
    - m_{\phi}^{2} }{m_X^{2} }
 \right)^2 \right] \nonumber
\end{eqnarray}

\begin{eqnarray*}
\mbox{where} ~~W^{3 \, \mu} = c_X^{} \, X^\mu \, + \cdots;\,\, 
\Big( W^{3 \, \mu} =  \sin \theta_w \, \tilde{A}^{ \mu}_{}  + 
\cos \zeta \cos \theta_w \, Z^{\mu}_- + \left( - \sin \zeta
 \cos \theta_w \right) Z^{\mu}_+ \Big)
\end{eqnarray*}

\subsubsection{Spin~0 dark matter}

The parameters associated with the initial bosons are labelled (i),
 and those with the final states labelled (f). The tree level cross
 sections for annihilation via a vector boson are given to leading
 order in $\beta_i^{}$, and are all
 p-wave annihilations:

\begin{itemize}
\item 
Dirac fermion anti-fermion product
\bigskip
\begin{eqnarray}
\sigma ~=~
P ~ \sqrt{1-\frac{m_f^2}{m_i^2}} ~
\left[ (c_V^2 + c_A^2)_f^X +  (c_V^2 - 2\, c_A^2)_f^X
 \left( \frac12 \frac{m_f^2}{m_i^2} \right)  \right] 
\label{ferm}
\end{eqnarray}

\begin{eqnarray}
\mbox{where} ~~~P ~=~ 
\frac{  g_X^4 \, (Q_X^{\, 2})_i }{(s-m_X^2)^2 + m_X^2 \Gamma_X^2} 
\left( \frac{ s \, \beta_i}{12 \pi}  \right)
\end{eqnarray}

\item
Spin~0, Spin~0 product
\begin{eqnarray}
\sigma &=&
P ~ (Q_X^{\, 2})_f ~ \frac14 \, \left( 1-\frac{m_f^2}{m_i^2} \right)^{3/2}
\end{eqnarray}

\item
 $ W^+ W^- $ product
\begin{eqnarray}
\sigma ~=~
P \,
\left( 1 -  \frac{m_W^2}{m_i^2}  \right)^{3/2}
\left(  \frac{g_W^2 \, c_X^2}{g_X^2}  \right)
\left( \frac{m_i }{ m_W } \right)^2  
\left[   1 + \frac34 \left( \frac{m_W }{ m_i } \right)^2 \right]
\end{eqnarray}

\item
Gauge boson ($X^\prime$) and a scalar ($\phi$) product
\vspace{1mm}
\begin{eqnarray}
\sigma &=&
P \,  \left( \frac{g_{X X^\prime }^2 }{g_X^2} \right) 
\sqrt{ 1-  \left( \frac{ m_{X^\prime}^{2} + m_{\phi}^{2} }{2 \,
    m_i^{2} }
 \right) +  \left( \frac{ m_{X^\prime}^{2} - 
m_{\phi}^{2} }{4 \, m_i^{2} } \right)^2 } 
\left[ \frac{v_\phi^2 }{32 \, m_{X^\prime}^2} \right]
 \nonumber \\
\nonumber \\
&& \hspace{10mm}
\times ~\left[ 
2 + \left( \frac{ 5 \, m_{X^\prime}^{2} - m_{\phi}^{2}  }{m_i^{2} } \right) 
+ \frac18 \left( \frac{ m_{X^\prime}^{2} - m_{\phi}^{2}  }{m_i^{2} } \right)^2 
  \right]
\end{eqnarray}

\end{itemize}

\begin{eqnarray}
\mbox{where}
\hspace{8mm}
v_H^2 &=& \frac{4 \, m_W^{2} }{g_W^{2} } \\
\nonumber \\
v_\phi^2 &=&  \left( \frac{m_{Z^\prime}^{} \,  \cos \chi  }{ g^{\prime
  }
 \, Q_{\phi}^{\prime } } \right)^2
\left[ 1 -   \frac{ \Delta^2 \, \cos^2 \theta_w^{} }{m_W^2 
 \left( m_{Z^\prime}^2 - m_W^2 \sec^2 \theta_w^{} \right)}  \right]   
\hspace{6mm}
\end{eqnarray}

\def\theequation{D-\arabic{equation}}
\def\thesubsection{D}
\setcounter{equation}{0}
\def\thefigure{D-\roman{figure}}
\setcounter{figure}{0}
\subsection{Sommerfeld effect}\label{appendixC}

In the following we provide  more
details about the Sommerfeld effect mentioned in Section~\ref{dmc}.
A more complete version of this study will appear elsewhere \cite{SC}.
When the initial or final state of an interaction event involves slow
 moving particles, attractive or repulsive forces between these
 particles can lead to large enhancements or suppressions in the cross
 section compared to a perturbatively calculated result. This has long
 been known in nuclear physics and condensed matter physics as the
 Sommerfeld effect. Recently, its importance has been pointed out for
 cosmological calculations of thermal relics, and the indirect
 detection of dark matter 
\cite{Sommerfeld,Hisano:2003ec,Hisano:2004ds,Hisano:2006nn,Cirelli:2007xd,
j1,MarchRussell:2008tu}.

The non-perturbative physics can be thought of as the limit of
 perturbative Feynman diagrams, with an infinite number of particle
 exchanges. For 2-body scattering this involves a ``ladder"
 diagram. The solution of the non-perturbative vertex function,
 $\Gamma$, is given by the Bethe-Salpeter equation:
\begin{figure*}[ht]
\center
\begin{minipage}{15.6cm}
\includegraphics[width=15.cm]{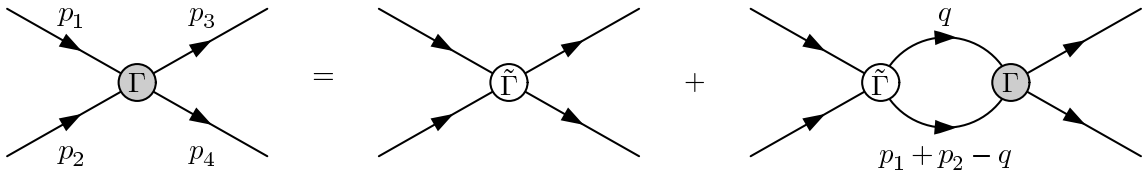} 
\end{minipage}
\caption{Bethe-Salpeter equation in diagrammatic form.}
\label{bs}
\end{figure*}

\vspace{-10mm}
\begin{eqnarray}
i \Gamma (p_1^{}, p_2^{}; p_3^{}, p_4^{} ) &=&
i \tilde{\Gamma} (p_1^{}, p_2^{}; p_3^{}, p_4^{} ) +
\int \frac{d^4 q}{(2 \pi)^4} ~ \tilde{\Gamma} ~ G(q) ~ G(p_1^{}
 + p_2^{} -q ) ~ \Gamma  
\end{eqnarray}
where $\tilde{\Gamma}$ are ``compact" vertices which do not involve
 any intermediate state composed solely of the scattering particles,
 and $G$ is the non-perturbative propagator. The inhomogeneous integral
 equation for $\Gamma$ can be solved to arbitrary accuracy, given
 that $\tilde{\Gamma}$ and the propagator are sufficiently well known
 from perturbative calculations. The finite ladders can be neglected
 when near the poles of the scattering states, which leads to the
 non-perturbative nature of this effect and a homogeneous integral
 equation. For an annihilation process as shown in Fig~\ref{annpic}, the
 matrix element for the diagram with the ladder is related to the one
 without a ladder as given below:
\begin{figure*}[h!]
\center
\includegraphics{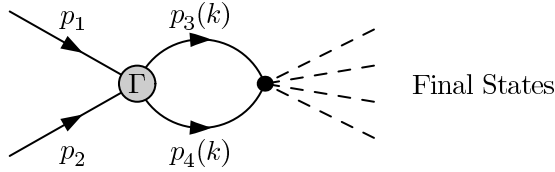} 
\caption{Diagram of Non-Perturbative Scattering before Annihilation}
\label{annpic}
\end{figure*}
\vspace{1mm}
\begin{eqnarray}
{\cal{M}}_{\mbox{\tiny with ladder}}^{} (p_1^{} , p_2^{} ; \{p_f^{}\}) &=& 
\int \frac{d^4 k}{(2 \pi)^4} ~{\cal{M}}_{\mbox{\tiny w/o}}^{} (p_3^{}
, p_4^{} ; \{p_f^{}\}) \, G(p_3^{}) \, G(p_4^{}) \,
\Gamma(p_1^{},p_2^{} ; p_3^{}, p_4^{})
 \nonumber \\
&=& 
\int \frac{d^3 k}{(2 \pi)^3} ~{\cal{M}}_{\mbox{\tiny w/o}}^{}
(\mathbf{k} ,
 - \mathbf{k} ; \{\mathbf{p}_f^{}\})
 ~\tilde{\psi}_{\mbox{\tiny BS}}^{} (\mathbf{k}),
\label{mladd}
\end{eqnarray}
where the Bethe-Salpeter wavefunction, $\tilde{\psi}_{\mbox{\tiny
 BS}}^{}$, is introduced in the centre of momentum frame, and the
 intermediate scattering particles are taken on-shell. It follows that
 for non-relativistic processes, the Bethe-Salpeter wavefunction is
 equivalent to the solution of the Schr\"{o}dinger equation with a
 potential that accounts for the scattering interactions 
\cite{Berestetsky:1982aq}.
For central potentials, the spherical harmonic basis is convenient
 for solving the Schr\"{o}dinger equation:
\begin{eqnarray}
\psi_{Elm}^{} (\mathbf{r}) &=&  R_{El}^{} (r) \,
 Y_{lm} (\theta_r^{} ,\phi_r^{} )
\label{sphh}
\end{eqnarray}
The Fourier transform of eq~(\ref{sphh}) is determined below:
\vspace{2mm}
\begin{eqnarray}
\tilde{\psi}_{Elm}^{} (\mathbf{k}) &=& \int d^3 r ~\psi_{Elm}^{}
 (\mathbf{r}) \, e^{i \mathbf{k \cdotp r}} \nonumber \\
&=&\!\!\!\!\! \int r^2 dr\!\! \int d \Omega_r^{} ~ R_{El}^{} (r) \, Y_{lm}
 (\theta_r^{} ,\phi_r^{} ) 
 \left[  \sum_{l^\prime=0}^\infty \sum_{m^\prime=-l^\prime}^{l^\prime}
 \!\!\! i^{l^\prime} 4 \pi \, j_{l^\prime}^{} (kr) \, Y_{l^\prime
 m^\prime}^{*}
 (\theta_r^{}, \phi_r^{}) \, Y_{l^\prime m^\prime}^{} (\theta_k^{},
 \phi_k^{})
 \right] \nonumber \\
&=& Y_{l m}^{} (\theta_k^{}, \phi_k^{}) \left[  i^{l}  \, 4 \pi 
 \int_0^\infty r^2 \, j_{l}^{} (kr) \, R_{El}^{} (r) \, dr \right] 
 \nonumber \\[9pt]
&\equiv& Y_{l m}^{} (\theta_k^{}, \phi_k^{}) ~ F_{El}^{} (k)
\end{eqnarray}
It is therefore useful to decompose the matrix element in the same
 orthonormal momentum space basis to calculate eq~(\ref{mladd}). The
 weighting of the components of the Bethe-Salpeter wavefunction with
 different ($l,m$) quantum numbers is determined by the matrix
 element. If ${\cal{M}}_{\mbox{\tiny w/o}}^{}$ is expressed as a
 polynomial in $k$, the result of integration of the wavefunction
 weighted by extra powers of $k$ needs to be found. This can be done
 by considering derivatives of the position space radial
 wavefunction and the inverse Fourier transform relation:
\begin{eqnarray}
R_{E l}^{} (r) &=& \frac{(-i)^l }{2\pi^2} \int_0^\infty k^2 j_l^{}
 (kr)
 F_{E l}^{} (k) ~dk  \\
\partial_r^{} R_{E l}^{} (r) &=& \frac{(-i)^l }{2\pi^2} \int_0^\infty
 k^2
 \partial_r^{} \left[ j_l^{} (kr) \right] F_{E l}^{} (k) ~dk
\end{eqnarray}
Using a series expansion of the spherical Bessel function, then
\begin{eqnarray}
\frac{\partial^{l+2n} R_{E l}^{}(r)}{\partial r^{l+2n}} \Bigg|_{r=0}
 &=& 
 \frac{ (l+2n)! \, (-1)^n}{2^n \, n! \, (2l+2n+1)!!} ~\frac{(-i)^l
 }{2\pi^2}
 \int_0^\infty k^{2} \, k^{l+2n} \, F_{E l}^{} (k)  \, dk
\label{derr}
\end{eqnarray}
Suppose the full matrix element is dominated by the contribution of
 one term in ${\cal{M}}_{\mbox{\tiny w/o}}^{} \propto k^{l+2n} \,
 Y_{lm}^{}  $, where n is some non-negative integer. In this case, the
 cross section which accounts for the ladder is related by an overall
 factor to the cross section calculated from diagrams without the
 ladder $(\sigma_{\mbox{\tiny with ladder}} = S_{l,n}^{} \,
 \sigma^{}_{\mbox{\tiny w/o}})$,
 valid in the non-relativistic limit. $S_{l,n}^{}$ is called the
 Sommerfeld factor, with the value:
\begin{eqnarray}
S_{l,n}^{} &=&  \left|  \left(  \frac{ n! \,\, 2^n  \, (2l+2n+1)!!}{
 (l+2n)! ~\, (M \beta)^{l+2n} }  \right)  \frac{\partial^{l+2n} R_{E
 l}^{}(r)}{\partial r^{l+2n}}
 \Bigg|_{r=0} \right|^2
\label{sfac}
\end{eqnarray}
It is common to ignore the effect present for non-zero angular
 momentum processes, arguing that the angular momentum barrier always
 suppresses higher partial wave processes. However, if s-wave
 interactions are small compared to higher partial wave
 processes, as when scalars annihilate via a heavy vector, this
 argument is not always applicable.

\subsubsection{Coulomb potential}

The reduced Schr\"{o}dinger radial equation for a two-body system
 of particles each with mass $M$, and a Coulomb potential is as follows:
\begin{eqnarray}
\left(\frac{\partial_r^{2}}{M} + M \beta^2 + \frac{A}{r}  -
 \frac{l(l+1)}{M r^2} \right) r \, R_l^{}(r \,; M, A, \beta) &=& 0 
\end{eqnarray}
where $\beta$ is the speed of each particle when at infinite
 separation in the centre of mass frame. This can be re-written in
 terms of dimensionless
 variables:
\begin{eqnarray}
 \left(   \partial_z^{2} + \frac14 + \frac{ x }{2 \, z}  -
 \frac{ l  \left( l+1\right)}{z^2} \right) z \, R_l^{} (z \, ; x) ~=~
 0
 \hspace{7mm} \mbox{where} \hspace{5mm}
z = 2 \, r M \beta \hspace{5mm}
x =  A / \beta 
\end{eqnarray}
The solutions are given in terms of a confluent hypergeometric function:
\begin{eqnarray}
R_{ l}^{}(z \, ;x) &=& e^{\pi x/4 }  ~ e^{- i z /2} \, z^l  \,
 ~\frac{\Gamma (1+\frac{ix}{2}+l) }{(2l+1)!}  \, \,   ^{}_1 F_{1}^{}
 \left( 1+\frac{ix}{2}+l \, ,  2l+2 \, ,  i z \right) \\
\nonumber \\
&=& e^{\pi x/4 } ~ e^{- i z /2} \, z^l ~ \sum_{j=0}^{\infty} ~\left[
 \frac{\Gamma \left( 1+\frac{ix}{2}+l +j \right)}{\left( 2l +1 +j
 \right)!} \,
 \frac{\left( i z \right)^j}{j !} \right]
\end{eqnarray}
The Sommerfeld factor written in this dimensionless formulation is:
\begin{eqnarray}
S_{l,0}^{}  &=&  \left|   \frac{ (2l+1)!}{ (l!)^2 }  \, 
 \frac{\partial^{l} R_{l}^{}(z)}{\partial z^{l}} \Bigg|_{z=0} \right|^2
\end{eqnarray}
The result for non-zero angular momentum states due to Coulomb
 interactions is thus:
\begin{eqnarray}
S_{l,0}^{} &=& S_{0,0}^{}
\times \prod_{b=1}^{l} \left( 1+  \frac{ x^2}{4  b^2}  \right)
\hspace{10mm} \mbox{where} \hspace{3mm} S_{0,0}^{} ~=~ 
\frac{\pi x}{1 - e^{ - \pi x } }
\end{eqnarray}
In a perturbative expansion, higher partial waves are suppressed by
 factors of $\beta^2$. However, in the limit of small velocities
 (large $x$) the Sommerfeld factor for a Coulomb interaction fixes
 each partial wave to have the same velocity
 dependence.

\subsubsection{Yukawa potential}

The Schr\"{o}dinger equation with a Yukawa potential is
 related to the Coulomb case as follows:
\begin{eqnarray}
A \to A \, e^{-m_X^{} r}  ~=~ A \,  e^{-x z /(2 y)}  
\hspace{5mm} \mbox{where} \hspace{3mm} y = A M / m_X^{}
\end{eqnarray}
where $m_X^{}$ is the mass of the mediator. The Sommerfeld factor for
 the $l=0$ case has been presented by Cirelli, Strumia and
 Tamburini~\cite{Cirelli:2007xd}. Figure~D-iii
 shows the $l=1$ case (n=0), where numerical simulations
 have been used to find the wavefunction.
\begin{figure}[ht]
\center
\subfloat[Attractive Potential]{\includegraphics[width=7cm] {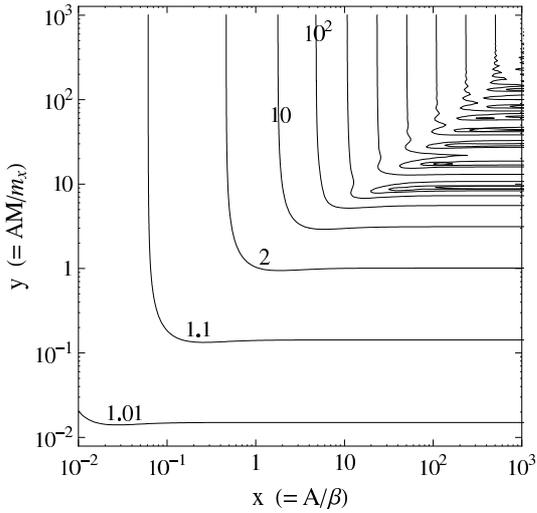}}
\hspace{10mm}
\subfloat[Repulsive Potential]{\includegraphics[width=7cm] {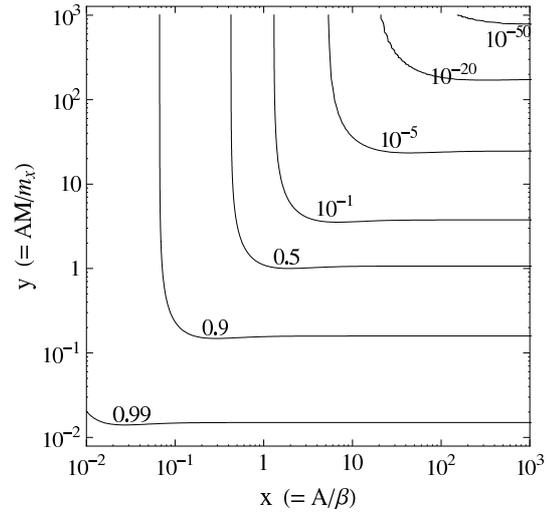}}
\label{sommtwo}
\caption{
Sommerfeld Factor for Yukawa interaction of a p-wave state 
(contours not labelled vary by a factor of 10)}
\end{figure}

For small $y$, the Sommerfeld factor gets closer to unity as the
 angular momentum increases. This is because as the interaction is
 screened beyond some short distance, the increased angular momentum
 barrier becomes more efficient at keeping the wavefunction away from
 this core. For a repulsive Yukawa interaction, there is then less
 suppression for the higher partial waves. So, although higher partial
 waves may be neglected at a perturbative level, this
 ladder effect could cause them to become significant or even
 dominant. However, for attractive interactions, the larger angular
 momentum processes are enhanced less, so higher partial wave terms
 which are negligible in the perturbative expansion
 stay negligible.

 There is always some reduction in the velocity dependent
 factor of the partial wave cross sections, in going to higher partial
 waves for non-zero $\beta$ , but the $\beta^{\, 2  \Delta l}$
 suppression is only found for large velocities. For the attractive
 Yukawa potential, $l=1$ bound states exist when $y \gtrsim 9.08$. The
 system can be close to the Breit-Wigner tails of these resonances
 when the relative velocity is small, and the Sommerfeld factor
 reflects this. In the limit
 $y \to \infty$, the Coulomb potential is approached, so the
 discussion in the previous subsection is relevant.

\end{document}